\documentclass[preprint,prd,tightenlines]{revtex4b5}
\usepackage{graphicx}
\usepackage{dcolumn}

\newcommand\gevcc{GeV/$c^2$}           %GeV/c^2
\newcommand\ttbar{$t\overline{t}$}     %ttbar
\newcommand\met{\mbox{$E\kern-0.57em\raise0.19ex\hbox{/}_{T}$}}
\newcommand\et{\mbox{$E_T$}}
\newcommand\half{\mbox{$\frac{1}{2}$}}
\newcommand\ppbar{$p\overline{p}$}

\begin{document}
\bibliographystyle{apsrev}

\preprint{Fermilab-Pub-01/057-E}

\title{
Search for First-Generation Scalar and Vector Leptoquarks}

% LIST_OF_AUTHORS.TEX                 5/18/01            
%
\author{                                                                      
%% names begin here                                                           
V.M.~Abazov,$^{23}$                                                           
B.~Abbott,$^{58}$                                                             
A.~Abdesselam,$^{11}$                                                         
M.~Abolins,$^{51}$                                                            
V.~Abramov,$^{26}$                                                            
B.S.~Acharya,$^{17}$                                                          
D.L.~Adams,$^{60}$                                                            
M.~Adams,$^{38}$                                                              
S.N.~Ahmed,$^{21}$                                                            
G.D.~Alexeev,$^{23}$
% ...
\\                                                                            
\vskip 0.30cm                                                                 
\centerline{(D\O\ Collaboration)}                                             
\vskip 0.30cm                                                                 
}                                                                             
\address{                                                                     
\centerline{$^{1}$Universidad de Buenos Aires, Buenos Aires, Argentina}       
\centerline{$^{2}$LAFEX, Centro Brasileiro de Pesquisas F{\'\i}sicas,         
                  Rio de Janeiro, Brazil}                                     
\centerline{$^{3}$Universidade do Estado do Rio de Janeiro,                   
                  Rio de Janeiro, Brazil}                                     
\centerline{$^{4}$Institute of High Energy Physics, Beijing,                  
                  People's Republic of China}                                 
\centerline{$^{5}$Universidad de los Andes, Bogot\'{a}, Colombia}             
\centerline{$^{6}$Charles University, Center for Particle Physics,            
                  Prague, Czech Republic}                                     
\centerline{$^{7}$Institute of Physics, Academy of Sciences, Center           
                  for Particle Physics, Prague, Czech Republic}               
\centerline{$^{8}$Universidad San Francisco de Quito, Quito, Ecuador}         
\centerline{$^{9}$Institut des Sciences Nucl\'eaires, IN2P3-CNRS,             
                  Universite de Grenoble 1, Grenoble, France}                 
\centerline{$^{10}$CPPM, IN2P3-CNRS, Universit\'e de la M\'editerran\'ee,     
                  Marseille, France}                                          
\centerline{$^{11}$Laboratoire de l'Acc\'el\'erateur Lin\'eaire,              
                  IN2P3-CNRS, Orsay, France}                                  
\centerline{$^{12}$LPNHE, Universit\'es Paris VI and VII, IN2P3-CNRS,         
                  Paris, France}                                              
\centerline{$^{13}$DAPNIA/Service de Physique des Particules, CEA, Saclay,    
                  France}                                                     
\centerline{$^{14}$Universit{\"a}t Mainz, Institut f{\"u}r Physik,            
                  Mainz, Germany}                                             
\centerline{$^{15}$Panjab University, Chandigarh, India}                      
\centerline{$^{16}$Delhi University, Delhi, India}                            
\centerline{$^{17}$Tata Institute of Fundamental Research, Mumbai, India}     
\centerline{$^{18}$Seoul National University, Seoul, Korea}                   
\centerline{$^{19}$CINVESTAV, Mexico City, Mexico}                            
\centerline{$^{20}$FOM-Institute NIKHEF and University of                     
                  Amsterdam/NIKHEF, Amsterdam, The Netherlands}               
\centerline{$^{21}$University of Nijmegen/NIKHEF, Nijmegen, The               
                  Netherlands}                                                
\centerline{$^{22}$Institute of Nuclear Physics, Krak\'ow, Poland}            
\centerline{$^{23}$Joint Institute for Nuclear Research, Dubna, Russia}       
\centerline{$^{24}$Institute for Theoretical and Experimental Physics,        
                   Moscow, Russia}                                            
\centerline{$^{25}$Moscow State University, Moscow, Russia}                   
\centerline{$^{26}$Institute for High Energy Physics, Protvino, Russia}       
\centerline{$^{27}$Lancaster University, Lancaster, United Kingdom}           
\centerline{$^{28}$Imperial College, London, United Kingdom}                  
\centerline{$^{29}$University of Arizona, Tucson, Arizona 85721}              
\centerline{$^{30}$Lawrence Berkeley National Laboratory and University of    
                  California, Berkeley, California 94720}                     
\centerline{$^{31}$University of California, Davis, California 95616}         
\centerline{$^{32}$California State University, Fresno, California 93740}     
\centerline{$^{33}$University of California, Irvine, California 92697}        
\centerline{$^{34}$University of California, Riverside, California 92521}     
\centerline{$^{35}$Florida State University, Tallahassee, Florida 32306}      
\centerline{$^{36}$University of Hawaii, Honolulu, Hawaii 96822}              
\centerline{$^{37}$Fermi National Accelerator Laboratory, Batavia,            
                   Illinois 60510}                                            
\centerline{$^{38}$University of Illinois at Chicago, Chicago,                
                   Illinois 60607}                                            
\centerline{$^{39}$Northern Illinois University, DeKalb, Illinois 60115}      
\centerline{$^{40}$Northwestern University, Evanston, Illinois 60208}         
\centerline{$^{41}$Indiana University, Bloomington, Indiana 47405}            
\centerline{$^{42}$University of Notre Dame, Notre Dame, Indiana 46556}       
\centerline{$^{43}$Iowa State University, Ames, Iowa 50011}                   
\centerline{$^{44}$University of Kansas, Lawrence, Kansas 66045}              
\centerline{$^{45}$Kansas State University, Manhattan, Kansas 66506}          
\centerline{$^{46}$Louisiana Tech University, Ruston, Louisiana 71272}        
\centerline{$^{47}$University of Maryland, College Park, Maryland 20742}      
\centerline{$^{48}$Boston University, Boston, Massachusetts 02215}            
\centerline{$^{49}$Northeastern University, Boston, Massachusetts 02115}      
\centerline{$^{50}$University of Michigan, Ann Arbor, Michigan 48109}         
\centerline{$^{51}$Michigan State University, East Lansing, Michigan 48824}   
\centerline{$^{52}$University of Nebraska, Lincoln, Nebraska 68588}           
\centerline{$^{53}$Columbia University, New York, New York 10027}             
\centerline{$^{54}$University of Rochester, Rochester, New York 14627}        
\centerline{$^{55}$State University of New York, Stony Brook,                 
                   New York 11794}                                            
\centerline{$^{56}$Brookhaven National Laboratory, Upton, New York 11973}     
\centerline{$^{57}$Langston University, Langston, Oklahoma 73050}             
\centerline{$^{58}$University of Oklahoma, Norman, Oklahoma 73019}            
\centerline{$^{59}$Brown University, Providence, Rhode Island 02912}          
\centerline{$^{60}$University of Texas, Arlington, Texas 76019}               
\centerline{$^{61}$Texas A\&M University, College Station, Texas 77843}       
\centerline{$^{62}$Rice University, Houston, Texas 77005}                     
\centerline{$^{63}$University of Virginia, Charlottesville, Virginia 22901}   
\centerline{$^{64}$University of Washington, Seattle, Washington 98195}       
}                                                                             
%end                                                                          

\begin{abstract}
We describe a search for the pair production of first-generation 
scalar and vector leptoquarks in the $eejj$ and $e\nu jj$ channels
by the D\O\ Collaboration.  The data are from the 1992--1996 \ppbar\ 
run at $\sqrt{s} = 1.8$ TeV at the Fermilab Tevatron collider.
We find no evidence for leptoquark production; in addition, no
kinematically interesting events are observed using relaxed selection
criteria.  The results from the $eejj$ and $e\nu jj$ channels 
are combined with those from a previous D\O\ analysis of the $\nu\nu jj$ 
channel to obtain 95\% confidence level (C.L.) upper limits on the leptoquark 
pair-production cross section as a function of mass and of $\beta$, the 
branching fraction to a charged lepton.  These limits are 
compared to next-to-leading-order theory to set 
95\% C.L. lower limits on the mass of a first-generation scalar leptoquark
of 225, 204, and 79~\gevcc\ for $\beta=1$, $\frac{1}{2}$, 
and 0, respectively.  For vector leptoquarks with gauge (Yang-Mills) couplings,
95\% C.L. lower limits of 345, 337, and 206~\gevcc\ are set on the mass for 
$\beta=1$, ${1 \over 2}$, and 0, respectively.  Mass limits for vector 
leptoquarks are also set for anomalous vector couplings.
\end{abstract}


\maketitle

\section{Introduction}

\subsection{Leptoquarks}

Leptoquarks (LQ) are exotic particles that couple to both leptons and quarks 
and carry color, fractional electric charge, and both lepton and baryon numbers
\cite{ann_rev}.
Although the pattern of three generations of doublets of quarks and leptons
suggests leptoquarks as a possible reason for an underlying unity, they
are not required in the standard model.
Leptoquarks, however, do appear in composite models, technicolor theories, 
grand unified theories, and superstring-inspired E$_6$ models.  They are not
part of the minimal supersymmetric (SUSY) standard model, but
can be accommodated in certain extended SUSY models. 
Leptoquarks can be scalar (spin 0) or vector (spin 1) particles.  In many 
models, both baryon and lepton numbers are conserved, allowing low-mass
leptoquarks to exist without mediating proton decay.

Leptoquarks with universal couplings to all flavors would
give rise to flavor-changing neutral currents and are severely
constrained by low-energy experiments.  We therefore assume in our analysis
that there is no intergenerational mixing and that, e.g., first-generation 
leptoquarks couple only to $e$ or $\nu_{e}$ and to $u$ or $d$ quarks.
In most models containing leptoquarks, each leptoquark species has a 
fixed branching fraction to $\ell^\pm q$: $\beta= 1$, \half\ or 0.  Models with 
intergenerational mixing or extra fermions can have any value of 
$\beta$ between 0 and 1.  

The H1 and ZEUS experiments at the $e^\pm p$ collider HERA at DESY have 
published lower limits on the mass of a first-generation leptoquark that 
depend on the unknown leptoquark-lepton-quark coupling, $\lambda$
\cite{h1_lq1,h1_lq2,h1_lq3,h1_lq4,h1_new,h1_newer,zeus_lq1,zeus_lfv,zeus_newer}.
Pair production of leptoquarks, nearly independent of the value of 
$\lambda$, could 
occur in $e^+ e^-$ collisions via a virtual $\gamma$ or $Z$ in the $s$-channel  
and in \ppbar\ collisions via an intermediary gluon.  Experiments at the CERN 
$e^+e^-$ LEP collider \cite{opal_lq,delphi_lq,l3_lq,aleph_lq} and at the 
Fermilab Tevatron \cite{d0_eejj,d0_enjj,cdf_new_lq} have searched for 
leptoquark pair production and have set lower limits on the masses of 
leptoquarks.

In February 1997, the H1 and ZEUS experiments reported an excess 
of events at high $Q^2$ \cite{h1_excess,zeus_excess}.  A possible 
interpretation of these events is the resonant production of first-generation 
leptoquarks at a mass ($M_{\text{LQ}}$) near 200 GeV/$c^2$ \cite{hewett}.  
Additional data collected in 1997 did not confirm this excess 
\cite{h1_new,zeus_new}.  (For a recent review of leptoquark phenomenology and 
the status of leptoquark searches at HERA and the Tevatron, see 
Ref. \cite{ann_rev}.)

\subsection{Leptoquark Production at the Tevatron}

At the Tevatron, pair production of leptoquarks can proceed through  
quark-antiquark annihilation (dominant for $M_{\text{LQ}} > 100$~\gevcc) 
and through gluon fusion,  
and is therefore independent of the LQ-$e$-$q$ Yukawa coupling
$\lambda$.  Pair production of first-generation leptoquarks can result in 
three final states: two electrons and two jets ($eejj$); one electron, a 
neutrino, and two jets ($e\nu jj$); or two neutrinos and two jets ($\nu\nu jj$).
The decay branching fractions
in the $eejj$, $e\nu jj$, and $\nu\nu jj$ channels are $\beta^2$,
$2\beta(1-\beta)$, and $(1-\beta)^2$, respectively.  The cross section for
$p\bar p \to \text{LQ}\;\overline{\text{LQ}} \to eejj$ 
is therefore proportional to $\beta^2$.  We use the next-to-leading-order (NLO) 
calculation of the pair-production cross section of scalar leptoquarks 
\cite{kraemer} to compare our experimental results with theory.  
This calculation has a theoretical uncertainty of about 15\% which corresponds 
to the variation of the renormalization scale $\mu$ used in the calculations 
between $\mu = 2M_{\text{LQ}}$ and $\mu = \half M_{\text{LQ}}$.  For vector 
leptoquarks, NLO calculations are not yet available, and we therefore use the 
leading-order (LO) pair-production cross section \cite{VSECT}.  
We consider three gluon 
couplings: Yang-Mills gauge couplings ($\kappa_{G} = \lambda_{G} = 0$),
minimal vector anomalous couplings ($\kappa_{G} = 1$ and $\lambda_{G} =0$),
and the anomalous couplings that yield the minimum cross section
for 150~\gevcc\ leptoquarks at $\sqrt{s} = 1.8$ TeV 
($\kappa_{G} = 1.3$ and $\lambda_{G} = -0.21$) \cite{VSECT}.

\section{D\O\ Detector and Triggering}
\label{sec:detector} 

The D\O\ detector is a general-purpose 
detector consisting of three major systems: a central tracking system, 
a uranium/liquid-argon calorimeter, and a muon 
spectrometer.  These are described in Ref. \cite{d0nim}.  
The features most relevant to this analysis are summarized below.

The central tracking system has a
cylindrical vertex drift chamber, a transition-radiation detector,
a cylindrical central drift chamber, and drift chambers in the forward regions.
The tracking system is used to determine the longitudinal ($z$) position of the 
\ppbar\ interaction and to find tracks associated with electrons and muons.  
Information from the transition-radiation detector helps separate electrons from
charged pions.  The calorimeter consists of a central calorimeter (CC) that 
covers the detector pseudorapidity \cite{eta_def} region 
$|\eta_{\text{det}}| < 1.2$ and two end calorimeters (EC) that cover 
$1.5 < |\eta_{\text{det}}| < 4.2$.
Scintillation counters located in the intercryostat region provide  
information about jets for $1.2 < |\eta_{\text{det}}| < 1.5$.
The electromagnetic (EM) and hadronic calorimeters
are segmented into cells in pseudorapidity and azimuthal angle 
($\phi$) of size $\Delta \eta_{\text{det}} \times \Delta \phi = 0.1 \times 0.1$ 
($0.05 \times 0.05$ at EM shower maximum). 

The Main Ring synchrotron lies above the Tevatron beam line and passes through 
the outer section of the central calorimeter.  Protons used for antiproton 
production pass through the Main Ring while the Tevatron is operating.  
Interactions in the Main Ring can cause spurious energy deposits in the 
calorimeter leading to false missing transverse energy (\met) in collected
events.  Certain triggers are rejected when the protons are being injected into
the Main Ring, every time the Main Ring beam passes through the detector, and 
during the subsequent ``calorimeter recovery'' period; other triggers are 
rejected during injection and when the proton bunch is present, but 
accepted during calorimeter recovery periods (called a ``minimal'' Main Ring 
veto).  Since all events are tagged with the state of the Main Ring at the time 
of collection, this rejection can be performed offline for triggers relying on
less restrictive Main Ring requirements.

D\O\ employs a three-level trigger system.  Level 0 uses
scintillation counters near the beam pipe to detect an inelastic
collision; Level 1 sums the EM energy in calorimeter
towers of size $\Delta \eta_{\text{det}} \times \Delta \phi = 0.2 \times 0.2$.  
Level 2 is a software trigger that forms clusters of calorimeter
cells and applies preliminary requirements on the shower shape.  Certain 
triggers also require energy clusters to be isolated.

\section{Event Reconstruction and Particle Identification}
\label{sec:id}

The D\O\ reconstruction program, {\footnotesize D\O RECO}, processes the 
triggered data into events with kinematic quantities and particle 
identification.  This includes finding interaction vertices, tracks, and 
jets, and identifying electrons and muons, each with loose quality criteria to 
reject poorly-measured objects.  Additional requirements are then applied for 
each analysis.

\subsection{Electron Identification}

Electron identification for the $eejj$ and $e\nu jj$ analyses is very similar.
Electron candidates are first identified by finding isolated
clusters of energy in the EM calorimeter.  These EM clusters are required to be 
in the fiducial volume of the detector, i.e.,  
$|\eta_{\text{det}}| < 1.1$ (CC) or 
$1.5 <|\eta_{\text{det}}| < 2.5$ (EC).  EM clusters with a matching
track from the primary vertex are called {\it electrons\/}; those without a
matching track are called {\it trackless electrons\/}.  A track and an EM
cluster in the CC match if the distance between the track and the EM cluster 
centroid is small:   
\begin{displaymath}
\sigma_{\text{trk}} = 
\sqrt{\left( \frac{\Delta\phi}{\delta_{\Delta\phi}} \right)^2 + 
      \left( \frac{\Delta z}{\delta_{\Delta z}} \right)^2} < 10,
\end{displaymath}
where $\Delta\phi$ is the azimuthal mismatch, $\Delta z$ is the mismatch along
the beam direction, and $\delta_x$ is the resolution for the observable $x$.
In the EC, $\Delta z$ is replaced by $\Delta r$, 
the mismatch transverse to the beam.

For the $eejj$ analysis, at least one of the two electrons in an event is
required to have a matching track.  An electron track can be improperly
reconstructed due to inefficiencies in the central tracking 
chambers or because of poor track/EM cluster matching caused by incorrect 
vertex information. Using trackless electrons restores some of this lost
efficiency, but at the expense of increased background.  They are not used in 
the $e\nu jj$ analysis.

For electron candidates with a matching track, we apply a likelihood test based 
on the following five variables:
\begin{itemize}

\item
Agreement between the observed shower shape and that expected for an 
electromagnetic shower.  This is computed using a 41-variable covariance matrix
for energy deposition in the cells of the electromagnetic 
calorimeter ($H$-matrix $\chi^2$~ \cite{d0_topsearch_prd}).

\item
The ratio of the shower energy found in the EM calorimeter to the total shower
energy, the electromagnetic energy fraction (EMF), is required to be that 
expected for an EM shower.

\item
A small track match significance, $\sigma_{\text{trk}}$, is required.

\item
The ionization $dE/dx$ along the track is required to be that for a single
minimum-ionizing particle.

\item
A variable characterizing the energy deposited in the transition-radiation
detector is required to be consistent with the expectation for an
electron
\end{itemize}
To a good approximation, these five quantities are independent of each
other for electron showers.
For EM objects without a matching track, an $H$-matrix $\chi^2 < 100$ is 
required.  

All EM objects are required to have deposited most of their energy in the
EM calorimeter ($\text{EMF} > 0.9$).  We also require EM objects to be isolated,
using the variable:
\begin{displaymath}
{\cal{I}} \equiv 
{{E_{\text{tot}}({\cal{R}}=0.4) - E_{\text{EM}}({\cal{R}}=0.2)} 
\over {E_{\text{EM}}({\cal{R}}=0.2) }}
\end{displaymath} 
where $E_{\text{tot}}({\cal{R}}=0.4)$ and $E_{\text{EM}}({\cal{R}}=0.2)$ are 
the total and EM energies in a cone of radius 
${\cal{R}} \equiv \sqrt{(\Delta\eta)^2 + (\Delta\phi)^2} = 0.4$ 
or 0.2 centered on the EM cluster, where the pseudorapidity is measured with
respect to the interaction vertex \cite{eta_def}.  
For electrons with matching tracks, we require ${\cal{I}} < 0.15$.
To reduce the multijet background by about 50\% in dielectron data in 
which one electron does not have a matching track, we require that electron to 
have ${\cal{I}} <0.10$.  The electron identification 
criteria are summarized in Table~\ref{table:elecid}.

\begin{table*}
\setlength\tabcolsep{8pt}
\caption{Electron identification requirements.}
\label{table:elecid}
\begin{ruledtabular}
\begin{tabular}{@{}l c c}
 Requirement & Electrons with tracks & Electrons without tracks \\
\hline
Fiducial volume             & $|\eta_{\text{det}} | < 1.1$ or 
                               $1.5 <|\eta_{\text{det}} | < 2.5$ &  
                              $|\eta_{\text{det}} | < 1.1$ or 
                               $1.5 <|\eta_{\text{det}} | < 2.5$ \\
Track match significance    & $\sigma_{\text{trk}}< 10$ \\
Electromagnetic fraction    & ${\text{EMF}} > 0.9$  & ${\text{EMF}} > 0.9$  \\
EM cluster isolation        & ${\cal{I}} < 0.15$ & ${\cal{I}} < 0.10$ \\
EM cluster shape $H$-matrix &              & $\chi^{2} < 100$ \\
5-variable likelihood       &    $< 1.0$   &                  \\
\end{tabular}
\end{ruledtabular}
\end{table*}

The electron $E_T$ resolution is $\sigma(E_T)/E_T = 0.0157 \oplus (0.072\ 
{\text{GeV}}^{1/2} / \sqrt{E_T}) \oplus 0.66\ {\text{GeV}}/E_T$, where $\oplus$ 
denotes a sum in
quadrature.  The resolution in $\eta$ and $\phi$ for an electron is
excellent, less than $10^{-2}$ \cite{d0_topmass_prd}.

\subsection{Jet Reconstruction}

Jet reconstruction \cite{jet_prd} is based on energy deposition in calorimeter 
towers (the calorimeter cells within 
$\Delta\eta \times \Delta\phi = 0.1 \times 0.1$) with 
$E_T > 1$~GeV.  Starting with the highest-$E_T$ tower, the
energy deposited in a cone of radius ${\cal{R}} = 0.7$ around the center
of the tower is summed and a new energy-weighted center is determined.  
This procedure is repeated, using the new center, until the jet's direction is 
stable.  Only jets with $E_T > 8$~GeV are retained.  The final direction of a 
jet is given by:
\begin{eqnarray*}
\theta_{\text{jet}} &=& \tan^{-1} \left[ 
{{\sqrt{(\sum_i E_x^i)^2 + (\sum_i E_y^i)^2}} \over {\sum_i E_z^i}}  \right] \\
\phi_{\text{jet}} &=& \tan^{-1} \left( {{\sum_i E_y^i} \over 
{\sum_i E_x^i}} \right) \\
\eta_{\text{jet}} &=& -\ln \left(\tan{{\theta_{\text{jet}}} \over {2}} \right)
\end{eqnarray*}
where the polar angle $\theta$ is measured relative to the interaction vertex, 
$E_x = E_i \sin(\theta_i) \cos(\phi_i)$, 
$E_y = E_i \sin(\theta_i) \sin(\phi_i)$, $E_z = E_i \cos(\theta_i)$, and $i$
corresponds to all cells that are within ${\cal{R}} = 0.7$.  Jets are 
required to have $|\eta_{\text{det}}| < 2.5$ and ${\text{EMF}} < 0.95$.

The measured jet energy is corrected for effects due to the underlying 
event and out-of-cone showering in the calorimeter. 
The transverse energy resolution for central jets ($|\eta_{\text{det}}| < 0.5$) 
varies from $\sigma(E_T)/E_T = 0.154$ for $E_T \approx 36$~GeV to 
$\sigma(E_T)/E_T = 0.050$ for $E_T \approx 300$~GeV \cite{jet_prd}.
The resolution in both $\eta$ and $\phi$ for 50 GeV jets varies from 
approximately 0.02 for $|\eta_{\text{det}}| < 0.5$ to approximately 0.06 for 
$2.0 < |\eta_{\text{det}}| < 2.5$ and improves as the jet energy
increases.

We use jets reconstructed with the large ${\cal{R}}=0.7$ cone size to 
decrease the number of final-state-radiation jets that are reconstructed 
separately from the parent jet and to improve the jet-energy and mass 
resolutions.  Jets are ordered in descending value of $E_T$, with $j_1$, the 
leading jet, having the highest $E_T$.

\subsection{Missing Transverse Energy}

Missing transverse energy is calculated as
\begin{displaymath}
\met = \sqrt{\met_x^2 + \met_y^2}
\end{displaymath}
where
\begin{eqnarray*}
\met_x &=& -\sum_i E_i\sin(\theta_i)\cos(\phi_i) - \sum_j \Delta E_x^j \\
\met_y &=& -\sum_i E_i\sin(\theta_i)\sin(\phi_i) - \sum_j \Delta E_y^j .
\end{eqnarray*}
The first sum is over all cells in the calorimeter and intercryostat detector
above the noise threshold,
and the second is over the corrections in $E_T$ applied to all electrons and
jets in the event.
The \met\ resolution is approximately 4 GeV per transverse component 
\cite{d0_dilep_topmass} and grows as the amount of calorimeter activity
increases.

\subsection{Vertex Finding}

The standard D\O\ vertex-finding algorithm uses tracks found in the central 
tracking system to locate the intersection of groups of tracks along the beam 
line.  The group with the largest number of tracks is chosen as
the primary vertex.  However, since there is an average of 1.5 interactions 
per beam crossing, the hard-scattering vertex is 
not always chosen correctly by this algorithm.  Using the electron to verify or 
recalculate the vertex significantly improves this efficiency \cite{W-mass}.  
The electron revertexing algorithm uses the track that best matches an 
EM calorimeter cluster and then recalculates the position of the vertex 
based on this track.  The $z$ position of the vertex is calculated by fitting a 
straight line through the centroids of the EM cluster and the matching 
track.  We require every event to contain at least one EM object with a 
matching track usable for revertexing.  If both EM clusters have a matching 
track, the primary vertex is calculated based on information from both 
of them.  The kinematic properties of the objects (electrons, jets, 
\met) in the event, such as transverse energy and 
pseudorapidity, are then recalculated based on the new vertex.  
All further analysis is done using the recalculated quantities.

\begin{figure}
\includegraphics[width=3.25in]{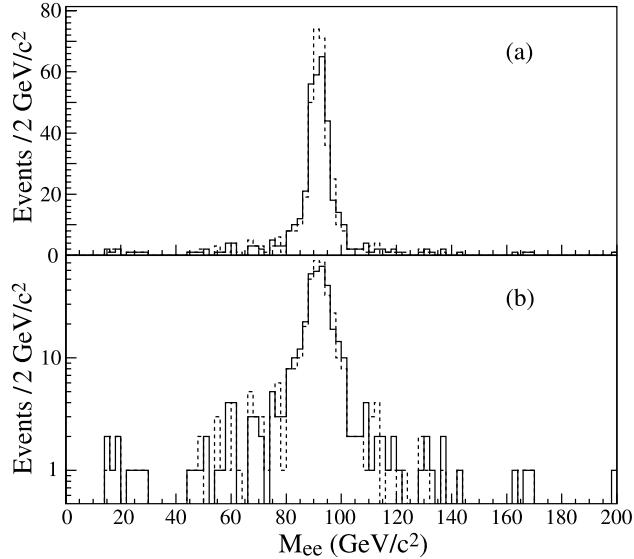}
\caption{$Z(\to ee)+2j$ data before (solid) and after (dashed) revertexing: 
(a) has a linear scale and illustrates the improvement in the $Z$-boson
mass resolution after the revertexing;  (b) has a 
logarithmic scale and shows the suppression of the background from 
vertex misidentification in the tails of the $Z$-boson peak.}
\label{fig:revertex}
\end{figure}

Figure~\ref{fig:revertex} illustrates the
improvement in the resolution of the $Z$-boson mass, as well 
as the reduction in background due to vertex misidentification for 
$Z(\to ee)+ 2j$ events, after the revertexing.  Events in this plot are allowed
to have one EM cluster without an associated track.  

\section{Search Strategies and Optimization}

The choice of variables, and the selection of their optimal values,
for improving the ratio of signal to background events is at the heart of 
searches for new particles.  We use two 
optimization techniques to aid in this selection: the random grid search 
method, which has been used by D\O\ in the measurement of the top-quark 
pair-production cross section \cite{d0_top_xsect} and in the search for the 
supersymmetric partner of the top quark \cite{susy_w_rgsearch}, and neural 
network analysis, which has been used by D\O\ in the measurement of the 
top-quark mass 
\cite{d0_topmass_prl,d0_topmass_prd} and in the determination of the 
\ttbar -to-all-jets cross section
\cite{top_alljets_xsect_prd,top_alljets_xsect_prl}.

\subsection{Additional Variables}
\label{sec:variables}

In addition to kinematic variables such as the transverse energies of 
electrons and jets and the \met\ used in 
standard analyses, we study other variables to determine their
efficiency in separating signal from background.  These include the energy 
sums, event-shape variables, invariant-mass variables, and mass-difference
variables listed below.

\begin{itemize}
\item Energy and transverse energy sums
\begin{eqnarray*}
H_T^e &-& {\text{sum of the \et\ of the two leptons}} \\  
 & & {\text{(two electrons, or electron and neutrino (\met))}} \\
H_T^j &-& {\text{sum of the \et\ of all jets}} \\
H_T^{j12} &-& {\text{sum of the \et\ of the two leading jets}} \\
H_T^{j123} &-& {\text{sum of the \et\ of the three leading jets}} \\
S_T &=& H_T^e + H_T^j \\
S_T^{12} &=& H_T^e + H_T^{j12} \\
S &-& {\text{total energy in the event}}
\end{eqnarray*}

\newpage

\item Event-shape variables
\begin{itemize}
\item[ ] centrality ($S_T/S$)
\item[ ] aplanarity of jets and leptons \cite{barger_phillips,d0_topmass_prd}
\item[ ] sphericity  \cite{barger_phillips} 
\item[ ] rms of the $E_T$-weighted distribution in jet $\eta$ 
\cite{top_alljets_xsect_prd}  
\end{itemize}

\item Invariant-mass variables
\begin{eqnarray*}
M_{ee} &-&  {\text{dielectron invariant mass}} \\
M_{ej} &-& {\text{invariant mass of various electron and jet }} \\
       & & {\text{combinations}} \\
M_T^{e\nu} &-&  {\text{electron-neutrino transverse mass}}
\end{eqnarray*}

\item Mass-difference variables for the $eejj$ analysis
\begin{eqnarray*}
{{\delta M}\over{M}}(M_{\text{LQ}}) &=& 
{\sqrt{(M_{\text{LQ}_1}-M_{{\text{LQ}}})^2 + 
(M_{\text{LQ}_2} - M_{{\text{LQ}}})^2}}\over{M_{\text{LQ}}} \\
{{\delta M}\over{M}} &=&
{M_{\text{LQ}_1}- M_{\text{LQ}_2}}
\over{(M_{\text{LQ}_1}+ M_{\text{LQ}_2})/2} \\
{\delta M}\over{\sqrt{M}} &=& {M_{\text{LQ}_1}- M_{\text{LQ}_2}} \over
{\sqrt{(M_{\text{LQ}_1}+ M_{\text{LQ}_2})/2}} 
\end{eqnarray*}

where $M_{\text{LQ}_1}$ and $M_{\text{LQ}_2}$ are the electron-jet
invariant-mass combinations that are closest to each other,  
and $M_{\text{LQ}}$ is the hypothesized leptoquark mass.

\item Mass-difference variable for the $e\nu jj$ analysis

\begin{displaymath}
{{\delta M}\over{M}}(M_{\text{LQ}}) = {\text{min}} \left( 
\frac {|M_{ej1}-M_{\text{LQ}}|}{M_{\text{LQ}}}, 
\frac {|M_{ej2}-M_{\text{LQ}}|}{M_{\text{LQ}}}\right)
\end{displaymath}
where $M_{ej1}$ and $M_{ej2}$ are the invariant masses of the electron with
the first and the second jet, respectively, and $M_{\text{LQ}}$ is the 
hypothesized leptoquark mass.  

\end{itemize}

Over fifty combinations of these variables were used in the random grid search 
and neural network studies described below to determine the optimal set of 
variables and selection criteria for the $eejj$ and $e\nu jj$ channels.

\subsection{Optimization Criterion}
\label{sec:optimization_criterion}

If first-generation leptoquarks with a mass of approximately 200~\gevcc\ exist, 
we want to achieve the highest-possible discovery significance.  If there is no 
evidence of leptoquark production, we want to set the lowest possible 95\% C.L. 
limit on their production cross section.
Based on the Monte Carlo (MC) simulations of the signal and the background
estimates described below, we pursue a fixed-background strategy for our
search.  We optimize our selection criteria by maximizing
the signal efficiency for 0.4 expected background events.  This method leads
to excellent discovery potential and a 67\% probability that no background
events will be observed.  If no events are observed, the experimental limit has
the advantage of being independent of the predicted number of background events 
and its uncertainty.

\subsection{Random Grid Search}

The random grid search method, which was implemented as the computer program 
{\footnotesize RGSEARCH} \cite{rgsearch}, helps determine the set of
cuts that optimally separates signal from background.
In a standard grid search, the signal and background acceptances for some 
cutoff ($x_{\text{cut}}$) on a variable $x$ are determined for all values 
between some minimum and maximum, $x_{\text{min}}$ and 
$x_{\text{max}}$, respectively.  A refinement of this technique is to use the 
MC signal to define the range of $x_{\text{cut}}$.  For
each MC event, $x_{\text{cut}}$ is set to the generated value of $x$, and the 
acceptances for signal and background are determined for that $x_{\text{cut}}$. 
While running {\footnotesize RGSEARCH}, the value of a cutoff on a 
variable can be fixed or allowed to vary in some range.  
Minimum and/or maximum values for $x_{\text{cut}}$ can be preset or, 
alternatively, any values that are allowed for signal can be used in the 
search.  In general, the search is multidimensional, and many combinations of 
variables, both fixed and varying, are studied to find an optimal set of
requirements to impose on the data.  Trigger thresholds and other criteria 
used to define the initial data sample are also imposed in all 
{\footnotesize RGSEARCH}  trials.  One of the results of an 
{\footnotesize RGSEARCH} trial is a plot of the number of expected signal 
events versus the predicted number of background events, normalized to the 
luminosity of the data sample, including detection efficiencies.  

\subsection{Neural Network Analysis}

We also use three-layer feed-forward neural networks (NN) \cite{jetnet,nn} in 
the search for leptoquarks.  For each combination of $n$ variables, a 
network is trained using MC signal events ($S$) and an appropriate mixture 
of background events ($B$) to yield an output discriminant 
${\cal D}_{\text{NN}}$ near 1 for signal and 0 for background.  For a
sufficiently large sample of training events, when the trained network is 
applied to the data, the discriminant output from the neural network is 
approximately $\frac{S(x)}{S(x)+B(x)}$, where $S(x)$ and $B(x)$ are the 
$n$-variable signal and background densities.  This defines contours of 
constant probability for signal versus background in the $n$-dimensional space 
that represent the optimal functions separating the signal from the background.
The discriminant then becomes a single variable that can be used to 
optimize the analysis for any desired signal to background ratio.

\section{The {\lowercase{\textit{eejj}}} Channel}

The study of the $eejj$ channel is particularly important because it
is the only channel sensitive to leptoquarks with $\beta = 1$.
It is also sensitive to leptoquarks with
$\beta < 1$; however, since both leptoquarks have to decay 
in the charged-lepton mode, the cross section for leptoquark pair production
and subsequent decay into the $eejj$ channel is suppressed by a factor of 
$\beta^2$.

Independent of the scalar or vector nature of leptoquarks, the analyses are very
similar.  In particular the data sample and the final event selection are 
identical.  We describe the scalar leptoquark analysis first, in detail, and 
then the vector leptoquark analysis.

\subsection{The Data}

\subsubsection{Triggers}

Events with two electrons satisfying the online trigger 
requirements listed in Table~\ref{table:trig} are used as the starting sample 
for the dielectron data sample.  
The total integrated luminosity for these triggers 
is~$123.0 \pm 7.0$~pb$^{-1}$, which corresponds to sample of 9519 events.
The average trigger efficiency for the data in this analysis 
is $(99.5 \pm 0.5)$\%.

\begin{table*}
\setlength\tabcolsep{8pt}
\caption{The Level 2 triggers used in the $eejj$ analysis.  The runs
listed correspond to different periods during Run 1 of the Tevatron
(1992--1996).  The transverse
energy of an EM cluster is denoted by $E_T^{\text{EM}}$.  The number of
events is that in the initial data set.} 
\label{table:trig}
\begin{ruledtabular}
\begin{tabular}{@{}c l c c}
Run & Trigger Requirements & Integrated Luminosity & Number of Events \\
\hline
Run 1A & $E_T^{\text{EM1,EM2}} > 10$ GeV & 14.7~pb$^{-1}$ & 1131 \\
Run 1B & $E_T^{\text{EM1}} > 20$ GeV, isolated  & 97.8~pb$^{-1}$ & 7500 \\
       & $E_T^{\text{EM2}} > 16$~GeV & & \\
Run 1C & $E_T^{\text{EM1}} > 20$ GeV, isolated & 10.5~pb$^{-1}$ & 888 \\
       & $E_T^{\text{EM2}} > 16$ GeV & & \\
\end{tabular}
\end{ruledtabular}
\end{table*}

\subsubsection{Event Selection for the Base Data Sample}

We require two electrons with $E_T^e > 20$~GeV and at least two jets with
$E_T^j > 15$~GeV.  
As described in Sec.~\ref{sec:id}, only one of the electrons is required to 
have a matching track.  Events containing an electron close to a jet
($\Delta {\cal{R}}_e < 0.7$) are rejected.     
Events whose dielectron invariant mass lies inside the $Z$-boson mass 
window, $ 82 < M_{ee} < 100$ GeV/$c^2$, are also removed.  
After identification, fiducial, initial kinematic, and $M_{ee}$ requirements, 
101 events remain.  We call these events the base data sample.

\subsection{MC Signal Samples}

Leptoquark pair production in the $eejj$ channel can be modeled as the 
production of a pair of identical strongly-interacting particles, each of 
which decays into an electron and a jet.
Monte Carlo events simulating the pair production of scalar leptoquarks 
are generated using {\footnotesize ISAJET} \cite{isajet} for leptoquark masses 
from 80 \gevcc\ to 250 \gevcc.  The {\footnotesize ISAJET} samples are used 
only for calculating acceptances; the NLO calculation of Ref.~\cite{kraemer} 
is used for the production cross section.

\subsection{Background Samples}
\label{sec:eejj_background}

The primary backgrounds to the $eejj$ final state are from $e^+e^-$ 
(``Drell-Yan'') production with
two or more jets, \ttbar\ production, and multijet events in which two 
jets are misidentified as electrons.

\subsubsection{Drell-Yan Background}

Drell-Yan (DY) events are generated using {\footnotesize ISAJET} in
four mass ranges: 20--60 \gevcc, 60--120 \gevcc, 120--250 \gevcc,
and 250--500 \gevcc.  For calculating the background, the DY$+2j$ 
cross section from {\footnotesize ISAJET} is normalized to the observed number 
of events in the $Z$-boson mass peak after imposition of the kinematic 
criteria described above.  
The scaling factor is $1.7\pm 0.1$ and reflects the fact that 
{\footnotesize ISAJET} does not provide the NLO corrections (``K-factor'') 
to the LO DY production cross section.  
The uncertainty in this background is 20\%, dominated by the 15\% uncertainty
in the jet energy scale.  We estimate that the base data sample contains 
$66.8 \pm 13.4$ DY events.

\subsubsection{\ttbar\ Background}

The \ttbar\ $\to$ dileptons MC sample is produced using {\footnotesize HERWIG} 
\cite{herwig} for $m_t = 170$ \gevcc.  The events are representative
of all $ee$, $e\mu$ and $\mu\mu$ final states, including those
from $\tau$ decay.  The sample of 101,339 events corresponds to an
integrated luminosity of about 270~fb$^{-1}$.  The D\O\ measurement 
\cite{d0_top_xsect} of the \ttbar\ production cross section has an uncertainty 
of 35\%.  This, when combined with the 15\% uncertainty in the jet energy 
scale, leads to an overall uncertainty of 38\% in the predicted number of 
\ttbar\ events.  The base data sample is estimated to contain $1.8 \pm 0.7$ 
\ttbar\ events.

\subsubsection{Photon Background}

Direct photon production is the main source of real photons (observed as
EM objects without associated tracks) in the $eejj$ final state; its 
contribution is small and is taken into account when the multijet 
background is estimated.  Other sources of photons, such as $W\gamma+2j$ 
production, are negligible for high-\et\ photons.

\subsubsection{Multijet Background}
\label{sec:QCD_misid}

The multijet background is estimated using data collected with 
a trigger that required three jets with $E_T^j > 10$~GeV at
Level 2.  This trigger was prescaled and had an integrated luminosity of
0.936~pb$^{-1}$.  Two sets of events are selected 
from this trigger.  Events in the $3j$ sample are required to have 
at least two jets with $E_T^j > 15$~GeV and at
least one additional jet with $E_T^j > 20$~GeV.  Events in the $2j+$EM sample 
have an EM object with $E_T^{\text{EM}} > 20$~GeV rather than a 
third jet. 

The probabilities for a jet to be misidentified as either an electron or 
trackless electron are determined by comparing the number of candidates with 
$E_T^e > E_0$ that pass standard quality cuts in the $2j+$EM 
sample and the total number of jets with $E_T^j > E_0$ in the $3j$ sample.  
The $E_0$ threshold is varied from 20 to 50 GeV, and the probabilities are 
stable for a cut value above 25~GeV, i.e., above the jet trigger turn-on.  The
probabilities for a jet to be misidentified as an electron with a track and
without a track are measured to be:
\begin{displaymath}
 f_{\text{track}} = (3.50 \pm 0.35)\times 10^{-4}; \quad
 f_{\text{no track}} = (1.25 \pm 0.13)\times 10^{-3},
\end{displaymath}
and, within the uncertainties, are independent of the $E_T$ and 
pseudorapidity of the electron.  These values are cross-checked using the ratio 
of $3j+$EM and $4j$ events.  This method of determining the misidentification 
probability automatically accounts for the direct photon background that is a 
part of the general ``multijet'' background.

We then apply these misidentification probabilities to the weighted number of
$4j$ events in the $3j$ sample.  The weight assigned to each event is the
number of jet permutations that can be used to misidentify a pair of EM
objects.  The backgrounds in the two samples, two electrons or an electron and
a trackless electron, are estimated by multiplying the weighted number of events
by $f_{\text{track}}^2$ or $2 f_{\text{track}} f_{\text{no track}}$,
respectively.  We assign an uncertainty of 15\% to these values, which reflects 
the variation of the misidentification probabilities as a function of $E_T^e$,
any difference between the CC and EC, as well as certain jet trigger turn-on 
effects.  The number of 
misidentified multijet events in the base data sample is estimated to be 
$24.3 \pm 3.6$ events.

\subsubsection{Total Background}

The total background estimate for the base data sample is $92.8 \pm 13.8$ 
events, in agreement with the 101 events observed in the data.

\subsection{Electron Identification Efficiencies}
\label{sec:e_eff}

There are approximately 300 $Z$-boson events remaining in the initial data 
sample after all requirements except those on the dielectron mass and for 
electron identification.  This is sufficient to estimate the
identification efficiencies for CC-CC, CC-EC, and EC-EC electron combinations.

We plot the dielectron mass spectrum without any electron identification 
requirements beyond EM object reconstruction and subtract the multijet and 
DY backgrounds using the standard ``side-band'' technique.  We then apply the 
electron identification requirements, again subtracting the backgrounds using 
the same side-band technique.  The ratio of the background-subtracted number of 
$Z$ bosons with the identification requirements to that without the 
identification requirements gives the efficiency per event.  The efficiency is 
($74\pm3$)\%, ($66\pm4$)\%, and ($68\pm9$)\% for CC-CC, CC-EC, and EC-EC 
electron combinations, respectively.

To calculate the average efficiency for leptoquark events, we find the
relative fractions of the CC-CC, CC-EC, and EC-EC topologies.  These are the
same, within the errors, for leptoquark masses of 180, 200, and 220~\gevcc,
and equal ($83 \pm 2$)\%, ($16 \pm 1$)\%, and ($1.1\pm 0.2$)\% for CC-CC, 
CC-EC, EC-EC combinations, respectively.  These fractions and the electron
identification efficiencies give an overall electron identification efficiency 
of ($73 \pm 4$)\% for leptoquark masses between 180 and 220~\gevcc.

\subsection{Event Selection Optimization}
\label{sec:eejj_rgsearch}

\subsubsection{Random Grid Search}

\begin{figure}
\includegraphics[width=3.25in]{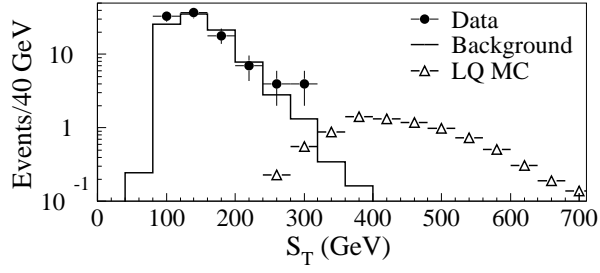}
\caption{$S_T$ distributions for background (solid line histogram),
data (solid circles), and $M_{\text{LQ}}=200$~\gevcc\ MC events (open triangles)
for the $eejj$ analysis.}
\label{fig:st}
\end{figure}

\begin{figure}
\includegraphics[width=2.75in]{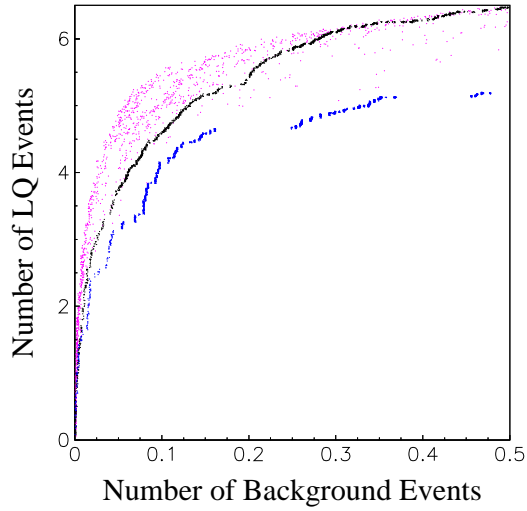}
\caption{Predicted number of $M_{\text{LQ}}=200$~\gevcc\ events versus the 
predicted number of background events for three {\footnotesize RGSEARCH} runs.  
The upper dotted line shows the variation with $S_T$.  
The lower dotted line shows the variation with $\frac{\delta M}{M}(200)$.
The structure (gaps) arises from an increase in acceptance for DY events.
The more dispersed set of dots shows the result when both 
$S_T$ and $\frac{\delta M}{M}(200)$ are varied.
The density of points is irrelevant.}
\label{fig:rgsearch_1}
\end{figure}

Extensive testing of combinations of the variables described in
Sec.~\ref{sec:variables} shows that the use of a single variable, the scalar 
sum of the transverse energies of all the objects in the event, $S_T$, is the 
most powerful.  Figure~\ref{fig:st} shows the $S_T$ distribution for the base 
data sample, the predicted background, and a sample of 200 GeV/$c^2$\ 
leptoquark MC events.  All of the leptoquark MC samples and the DY, \ttbar, and 
$2j+$EM background samples are used in the random grid search.  The leptoquark 
events are used to set the trial threshold values for the different 
parameters.  The number of predicted background events is determined
using the three background samples.  Shown in Fig.~\ref{fig:rgsearch_1} is the 
predicted 
number of signal events versus the expected number of background events for 
three different {\footnotesize RGSEARCH} trials, where the samples have been 
normalized to an integrated luminosity of 123 pb$^{-1}$, and the detection 
efficiencies, as well as the kinematic acceptance for the 
{\footnotesize RGSEARCH} thresholds, have been included.  In these trials, the 
$E_T$ thresholds of
the two electrons and the two jets are fixed to those in the base data sample.
The thresholds varied are those for $S_T$ alone, for 
${{\delta M}\over{M}}(200)$ alone,
and for these two variables together.  When combined,
$S_T$ and the mass-difference variable yield a higher signal efficiency
for very low values of expected background (less than 0.3 events), but 
the result is comparable to the use of $S_T$ alone when the expected background 
is approximately 0.4 events.  For the same expected background, using just the 
mass-difference variable leads to a 10\% reduction in the predicted number of 
signal events compared to that using just $S_T$.  Requiring $S_T > 350$ GeV 
leads to approximately 0.4 expected background events (see
Sec.~\ref{sec:optimization_criterion}).  The highest value 
of $S_T$\ seen in the data is 312~GeV, therefore no events pass 
this requirement.

\subsubsection{Neural Network Analysis}

\begin{figure}
\includegraphics[width=3.25in]{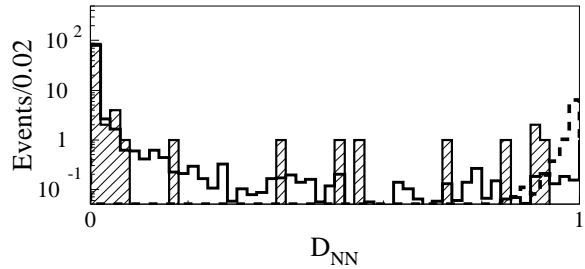}
\caption{Comparison of ${\cal{D}}_{\text{NN}}$ distributions for the predicted 
background (solid line histogram), 200~\gevcc\ leptoquark events (dashed line 
histogram), and the data (hatched histogram).}
\label{fig:dnn_eejj}
\end{figure}

The analysis based on the random grid search uses the linear sum 
$S_T \equiv H_T^e + H_T^j$.  
However, it is possible that a function other than a simple linear sum is the 
optimal way to combine the two variables.  The simplest way to compute this 
function is with a two-dimensional neural network.
For this approach, we use a neural network with two input nodes 
(corresponding to the variables $H_T^e$ and $H_T^{j}$), three hidden nodes,
and one output node.  The network is trained using the 200~GeV/$c^2$ 
leptoquark MC sample as signal (with a desired network output 
${\cal D}_{\text{NN}} = 1$) and the observed admixture of DY, \ttbar, and
multijet events as background (with desired ${\cal D}_{\text{NN}} = 0$). 
Figure~\ref{fig:dnn_eejj} shows the distribution of ${\cal D}_{\text{NN}}$ for 
the background, the 200~\gevcc\ leptoquark MC events, and the data.  The 
discrimination between signal and background is good. 

\begin{figure}
\includegraphics[width=3.35in]{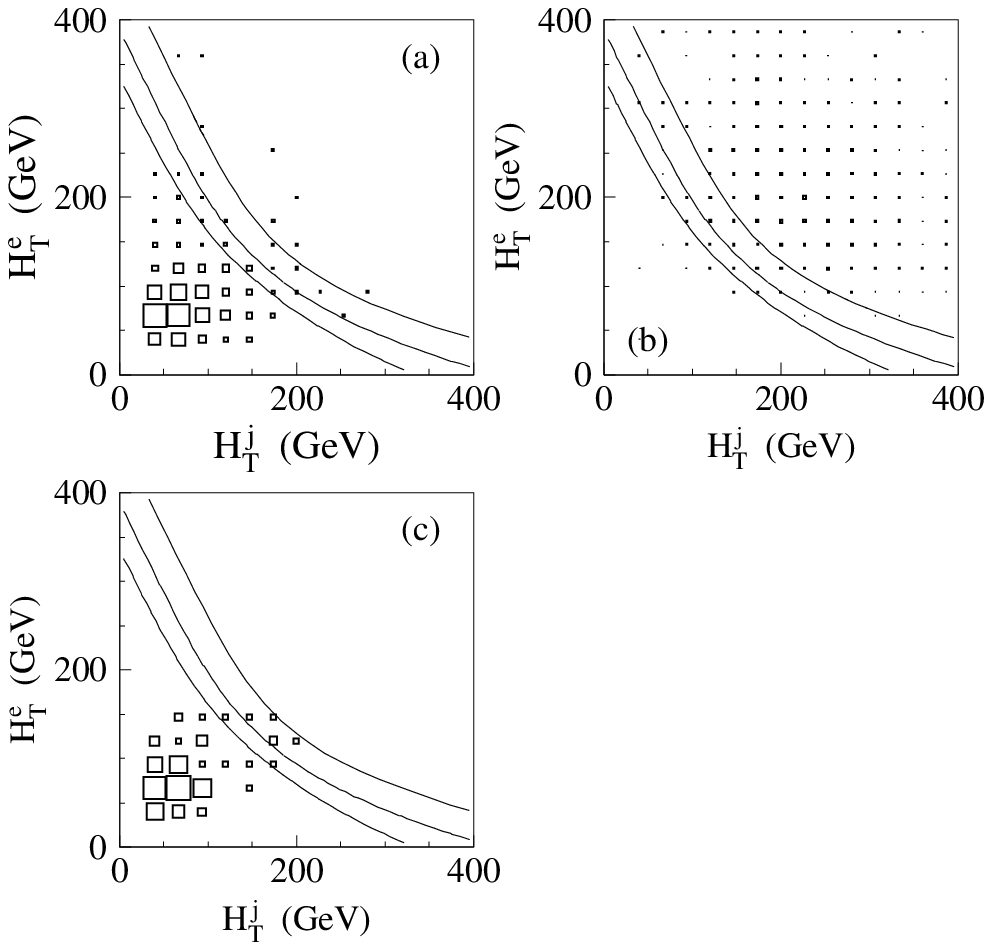}
\caption{$H_T^e$ versus $H_T^j$ for (a) the predicted background,
(b) 200~\gevcc\ leptoquark events, and (c) the base data sample.  The curved 
lines correspond to ${\cal{D}}_{\text{NN}} = 0.5$, 0.8, and 0.95 (from left to 
right).  The area of a displayed square is proportional to the number of events 
in that bin, with the total number of events normalized to 123 pb$^{-1}$.}
\label{fig:nn_eejj}
\end{figure}

Each value of ${\cal D}_{\text{NN}}$ defines a contour of constant probability 
between signal and background in the $(H_T^e, H_T^j)$ plane.  
The expected distributions in $x \equiv (H_T^e, H_T^j)$ space for a 200~\gevcc\ 
leptoquark signal, the background,
and the data are shown in Fig.~\ref{fig:nn_eejj}.  The contours 
corresponding to ${\cal D}_{\text{NN}} = 0.5$, 0.8, and 0.95 are
also shown.
 
Selecting events with ${\cal{D}}_{\text{NN}} > 0.95$ yields approximately 0.4 
background events.  The highest value of ${\cal D}_{\text{NN}}$ in the 
data is $0.92$ and no events survive the selection.  The efficiency for
identifying 200~\gevcc\ leptoquark events using the NN analysis is nearly 
identical to the efficiency found using the $S_T$ analysis.
Since the two methods give essentially equivalent results for the 
final experimental limits, we use the simpler $S_T$ analysis based on the 
random grid search described in Sec.~\ref{sec:eejj_rgsearch}.

\subsection{Checks}

\subsubsection{$S_T$ Distribution}
\label{sec:Z-peak}

\begin{figure}
\includegraphics[width=3.25in]{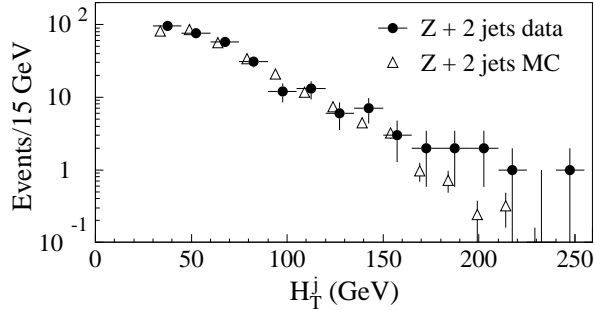}
\caption{The $H_T^j$ distribution for $Z+2j$ data (solid circles) and MC (open
triangles) in the $Z$\-boson mass region.  For high-mass DY events, 
$S_T \approx 350$~GeV corresponds to $H_T^j \approx 100$~GeV.}
\label{fig:htj_Z_2j}
\end{figure}

The modeling of the $S_T$ distribution for high-mass
DY events is checked by studying $H_T^e$ and $H_T^j$ separately, using
data and MC events in the $Z$-boson mass region.
The average value of $H_T^e$ for high-mass DY events (which provide most
of the DY background) is approximately
250 GeV, corresponding to an $H_T^j$ of approximately 100 GeV for
$S_T = 350$ GeV.  The distribution of $H_T^j$ for high-mass DY events is
expected to be similar to that for $Z+2j$ events.  Figure~\ref{fig:htj_Z_2j}
shows the $H_T^j$ distribution for $Z+2j$ MC and data.  In the region
corresponding to the $S_T$ cutoff for high-mass DY events ($H_T^j \approx
100$~GeV), the agreement is good.  Disagreement between the $Z+2j$ MC events
and the data at higher values of $S_T$ stems from the LO calculations used in 
the simulation and does not affect the results of this analysis.

\begin{figure}
\includegraphics[width=2.75in]{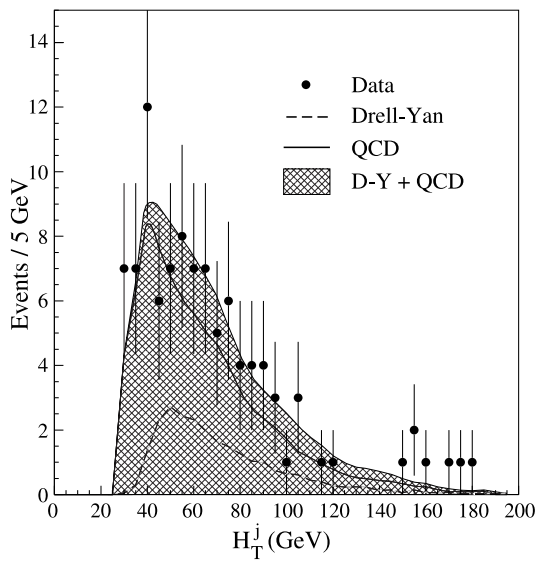}
\caption{Fit of the $H_T^j$ distribution in the $eejj$ data to the sum
of the DY and multijet backgrounds.}
\label{fig:lqfit}
\end{figure}

In addition, we fit the $H_T^j$ distribution of the data to a sum of the DY and
multijet backgrounds (the expected \ttbar\ background is smaller
than the uncertainties in the fit and is neglected).    
Figure~\ref{fig:lqfit} shows the $H_T^j$ distribution for the data and the
result of the fit for the two backgrounds.   The fit  
yields $77.5 \pm 15.9$ DY events and $24.6 \pm 13.9$ misidentified multijet
events, for a total of $102 \pm 21$ events, in agreement with the 101
events in the base data sample and with the direct determination of the two
dominant background contributions.

\subsubsection{Mass Fitting}

To improve resolution, rather than simply calculating the invariant masses of 
the electron-jet pairs, we use a kinematic fitter to reconstruct the mass of 
two identical particles that decay to electron+jet. 
The D\O\ fitting package, {\footnotesize KFIT}, is based on the bubble-chamber 
fitting program {\footnotesize SQUAW} \cite{squaw}.

The fitter balances the two electrons and the two leading jets against any 
extra jets and unclustered energy in the event by minimizing a $\chi^2$ to find 
the best fit solution.  The $\chi^2$ takes into account the object resolutions
(see Sec.~\ref{sec:id}) as well as the kinematic constraints.
Three constraints are used in the fit: momentum conservation in the 
$x$ and $y$ directions for electrons, jets and unclustered energy, and 
the equivalence in the mass of the two leptoquarks.

In each event there are two ways to associate the electrons and two leading 
jets ($e_1j_1,e_2j_2$ and $e_1j_2,e_2j_1$).  Fits for both configurations are
performed and the configuration with the lowest $\chi^2$ is retained.
The mass distribution for the background is found using the MC samples for DY
and \ttbar\ events; the multijet sample is not large enough to parametrize
a smooth line shape, so a jet is used to simulate an
electron in the fit.  

\begin{figure}
\includegraphics[width=3.35in]{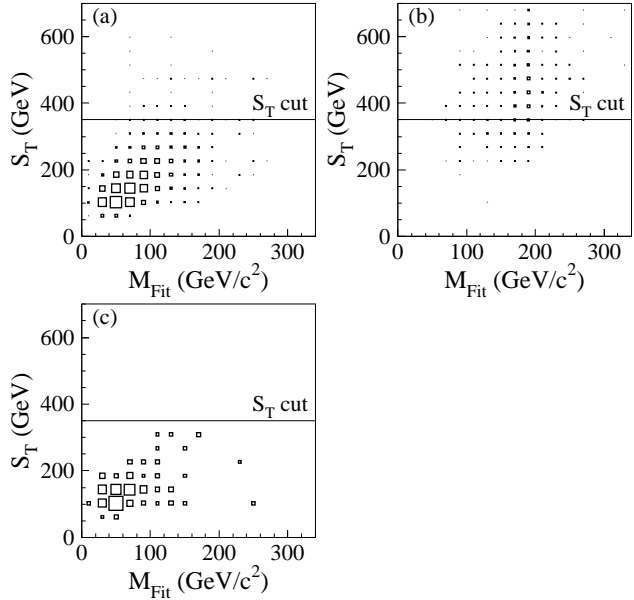}
\caption{$S_T$ versus the fitted mass for (a) background, (b) 200~\gevcc\ 
leptoquarks, and (c) the base data sample.  The area of a displayed square 
is proportional to the number of events in the bin. }
\label{fig:st_mfit_eejj}
\end{figure}

\begin{figure}
\includegraphics[width=3.4in]{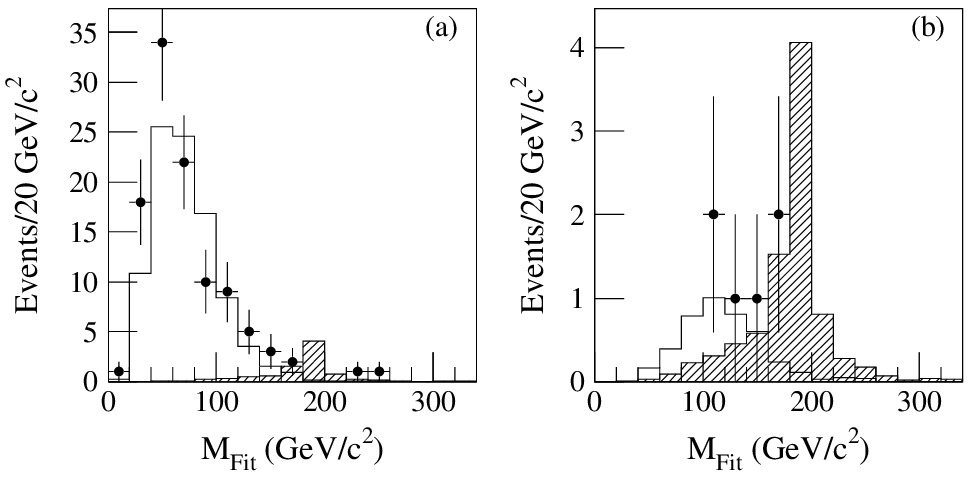}
\caption{Distributions of fitted mass for the events in the base 
data sample (solid circles), expected background (solid line histogram), and
200~\gevcc\ leptoquarks (hatched histogram) with (a) no cut on $S_T$ and 
(b) a reduced threshold of $S_T > 250$~GeV.}
\label{fig:mfit_eejj}
\end{figure}

Figure~\ref{fig:st_mfit_eejj} shows $S_T$ as a function of the fitted mass 
for the background, the 200~\gevcc\ leptoquark MC sample, and the data,
before the $S_T > 350$~GeV requirement.  The background is centered at 
low $S_T$ and low fitted mass and does not resemble the leptoquark signal.
The data most closely resemble the 
expected background.  Figure~\ref{fig:mfit_eejj} displays the one-dimensional 
distributions in fitted mass for the three samples before the $S_T$ cut and 
with a reduced $S_T > 250$~GeV requirement.  The data and the predicted 
background are in good agreement. 

\subsubsection{Varying the $S_T$ Threshold}

Table~\ref{table:bkg_ST} shows a comparison between the predicted number of
events from each of the three background sources, the total background, and the
number of events observed in the data as a function of $S_T$ threshold.  
The agreement between the predicted background and the data is excellent.

\begin{table}
\setlength\tabcolsep{3pt}
\caption{Comparison of the number of events expected from background with the
number observed for the $eejj$ analysis as a function of the threshold on 
$S_T$.}
\label{table:bkg_ST}
\begin{ruledtabular}
\begin{tabular}{@{}cccccc}
$S_T$ Threshold & DY & Multijet & \ttbar & Total Background & Data \\
 (GeV) & & & & & \\ 
\hline
  0 & 66.8 & 24.3 & 1.79 &  $92.8 \pm 13.8$  & 101 \\
100 & 61.0 & 23.2 & 1.79 &  $85.9 \pm 12.7$  &  85 \\
125 & 45.0 & 16.9 & 1.75 &  $63.7 \pm  9.36$ &  63 \\
150 & 28.8 & 10.2 & 1.65 &  $40.6 \pm  5.96$ &  39 \\
175 & 16.0 & 5.67 & 1.44 &  $23.1 \pm  3.32$ &  20 \\
200 & 9.12 & 3.16 & 1.15 &  $13.4 \pm  1.93$ &  15 \\
225 & 4.88 & 1.73 & 0.84 &  $7.45 \pm  1.06$ &  9 \\
250 & 2.64 & 0.99 & 0.59 &  $4.22 \pm  0.59$ &  8 \\
275 & 1.35 & 0.60 & 0.39 &  $2.34 \pm  0.32$ &  5 \\
300 & 0.75 & 0.35 & 0.25 &  $1.35 \pm  0.19$ &  3 \\
325 & 0.31 & 0.23 & 0.16 &  $0.70 \pm  0.09$ &   0 \\
350 & 0.18 & 0.16 & 0.11 &  $0.44 \pm  0.06$ &   0 \\
375 & 0.12 & 0.11 & 0.07 &  $0.30 \pm  0.04$ &   0 \\
400 & 0.07 & 0.08 & 0.04 &  $0.20 \pm  0.03$ &   0 \\
\end{tabular}
\end{ruledtabular}
\end{table}

\subsection{Signal Studies}

\subsubsection{Systematic Uncertainties}

The systematic uncertainties in the signal acceptance are obtained by 
comparing the results for scalar leptoquark samples 
generated using {\footnotesize ISAJET} and {\footnotesize PYTHIA} with 
different structure functions and renormalization scales.  The uncertainty from 
the jet energy scale is determined by varying the calorimeter response to 
jets by one standard deviation.  The systematic error in the signal varies 
from 17\% to 13\% for leptoquark masses between 120 and 250~\gevcc.  The 
systematic uncertainties are summarized in Table~\ref{table:syst}.

\begin{table}
\caption{Systematic uncertainties in the signal for the $eejj$ analysis.}
\label{table:syst}
\begin{ruledtabular}
\begin{tabular}{@{}ll}
Source & Uncertainty (\%)\\
\hline
Particle identification & 5 \\
Smearing in the detector & 3 \\
Jet energy scale & 11--2 ($M_{\text{LQ}} =$ 120--250 \gevcc) \\
Gluon radiation & 7 \\
PDF and $Q^2$ scale & 7 \\
Monte Carlo statistics & 2 \\
Luminosity & 5 \\
\hline
Total & 17--13 ($M_{\text{LQ}} =$ 120--250 \gevcc) \\
\end{tabular}
\end{ruledtabular}
\end{table}

\subsubsection{Signal Efficiency}

The signal-detection efficiencies are determined using simulated scalar 
leptoquark events that pass the selection requirements and are shown in 
Table~\ref{table:cs-ee}.  The uncertainties in the efficiencies include 
uncertainties in trigger and particle identification, the
jet energy scale, effects of gluon radiation and parton fragmentation in
the modeling, and finite Monte Carlo statistics.  The overall efficiency
ranges from 1\% to 38.5\% for leptoquark masses between 80~\gevcc\ and 
250~\gevcc.

\begin{table}
\setlength\tabcolsep{6pt}
\caption{Efficiency, background, 95\% C.L. upper limit on the leptoquark pair
production cross section ($\sigma_{\text{limit}}$), and the NLO cross section 
($\sigma_{\text{NLO}}$) with $\mu=2M_{\text{LQ}}$
\protect \cite{kraemer} for $\beta=1$ as a function of 
leptoquark mass for the $eejj$ channel.}
\label{table:cs-ee}
\begin{ruledtabular}
\begin{tabular}{@{}ccccc}
Mass & Efficiency & Background & $\sigma_{\text{limit}}$ & 
                                                  $\sigma_{\text{NLO}}$ \\
(\gevcc) & (\%) & (Events) & (pb) & (pb) \\ 
\hline
 80 & $ 1.0 \pm 0.2$ & $0.44 \pm 0.06$ & 2.9   & 36.0 \\
100 & $ 3.4 \pm 0.6$ & $0.44 \pm 0.06$ & 0.80  & 10.7 \\
120 & $ 8.8 \pm 1.4$ & $0.44 \pm 0.06$ & 0.30  & 3.81 \\
140 & $14.4 \pm 2.1$ & $0.44 \pm 0.06$ & 0.18  & 1.54 \\
160 & $20.9 \pm 3.0$ & $0.44 \pm 0.06$ & 0.13  & 0.68 \\
180 & $27.6 \pm 3.8$ & $0.44 \pm 0.06$ & 0.094 & 0.32 \\
200 & $33.2 \pm 4.0$ & $0.44 \pm 0.06$ & 0.076 & 0.16 \\ 
220 & $36.1 \pm 4.4$ & $0.44 \pm 0.06$ & 0.070 & 0.080 \\
225 & $37.7 \pm 4.5$ & $0.44 \pm 0.06$ & 0.067 & 0.068 \\
250 & $38.5 \pm 4.7$ & $0.44 \pm 0.06$ & 0.066 & 0.030 \\
\end{tabular}
\end{ruledtabular}
\end{table}

\subsection{Results from the {\mbox{$eejj$}} Channel for Scalar Leptoquarks}

\begin{figure}
\includegraphics[width=3.0in]{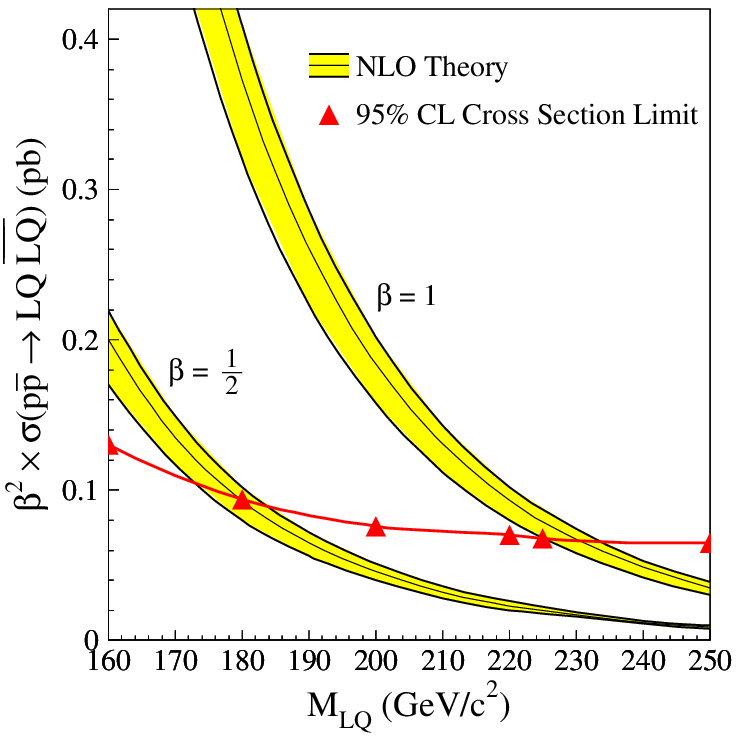}
\caption{Upper limit on the leptoquark pair-production cross section 
(triangles) from the $eejj$ channel.
The NLO calculations of Ref.~\protect \cite{kraemer} for $\beta=1$ (upper band)
and $\beta=\half$ (lower band) are also shown.  The central lines correspond to 
$\mu = M_{\text{LQ}}$, and the lower and upper edges of the bands correspond to
$\mu = 2M_{\text{LQ}}$ and $\mu = \half M_{\text{LQ}}$, respectively.}
\label{fig:slqlim_eejj}
\end{figure}

Based on our observation of no events after requiring $S_T >
350$~GeV, we set a 95\% C.L. upper limit on the leptoquark pair-production
cross section using a Bayesian approach \cite{limit_setting}
with a flat prior distribution for the signal cross section.  Limits for 
different leptoquark masses are summarized in Table~\ref{table:cs-ee}.
As indicated before, to compare our experimental results with theory, we use 
the NLO calculation of the production cross section \cite{kraemer}.  This cross 
section is tabulated for a wide range of leptoquark masses and has the value of 
$0.184_{-0.026}^{+0.018}$~pb for a 200~\gevcc\ leptoquark.  The theoretical 
uncertainty corresponds to
variation of the renormalization scale $\mu$ used in the calculation from
$2M_{\text{LQ}}$ to $\half M_{\text{LQ}}$.  To set a limit on the leptoquark 
mass, we compare the theoretical cross section for
$\mu = 2M_{\text{LQ}}$ with our experimental limit, resulting in 
$M_{\text{LQ}} > 225$~\gevcc\ for a scalar leptoquark with $\beta=1$ and 
$M_{\text{LQ}} > 176$~\gevcc\ for a scalar leptoquark with $\beta=\half$. 
Figure~\ref{fig:slqlim_eejj}
shows the experimental limit as a function of scalar leptoquark
mass along with the predicted cross sections for $\beta=1$ and
$\beta=\half$.  The CDF collaboration has set a lower limit of 
$M_{\text LQ} > 213$~\gevcc\ \cite{cdf_new_lq} for $\beta=1$.  When our result
is combined with the CDF limit, a Tevatron mass limit of 
$M_{\text{LQ}} > 242$~\gevcc\ is obtained for $\beta = 1$ \cite{comb_lq}.

\subsection{Vector Leptoquarks}

Vector leptoquark events were generated for 
leptoquark masses from 100 \gevcc\ to 425 \gevcc\ using a version of 
{\footnotesize PYTHIA} \cite{pythia} modified to include vector leptoquarks 
with various couplings. 
%\cite{msu}.  
The distributions of the kinematic variables for scalar and vector leptoquarks 
are sufficiently similar that the same event selection can be used for both 
analyses. 

\begin{figure}
\includegraphics[width=3.0in]{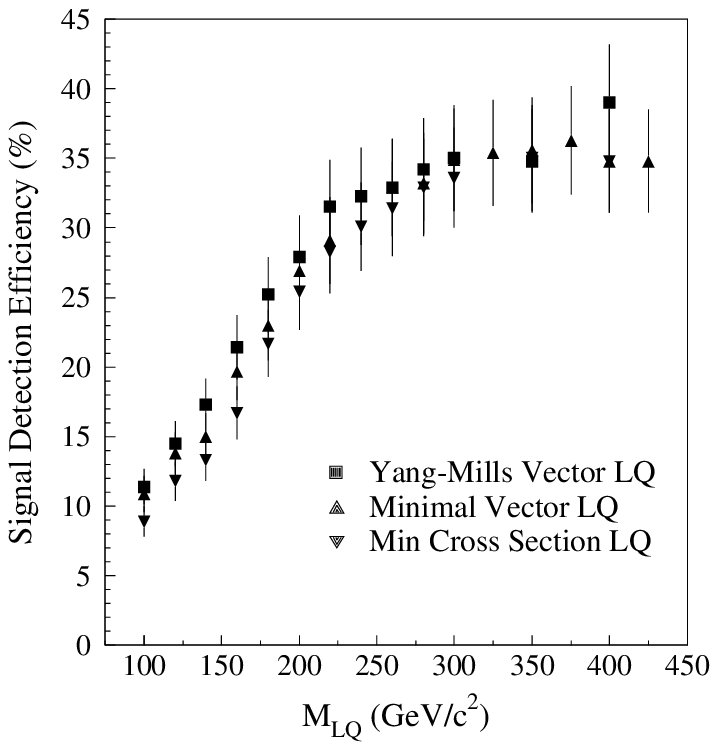}
\caption{The efficiency for identifying vector leptoquarks for the three
couplings in the $eejj$ channel.  The differences between the efficiencies are
small relative to the uncertainties.}
\label{fig:eeeff}
\end{figure}

The identification efficiencies for vector leptoquarks for the three couplings
considered are identical within their uncertainties, as shown in 
Fig.~\ref{fig:eeeff}.  To reduce the statistical uncertainty from the MC,
we use the average identification efficiency of the three sets of MC
events to set a single experimental limit on the cross section.  This limit is 
then compared with the appropriate prediction for each coupling.

The cross sections for vector leptoquark production have been 
calculated only to LO for three gluon couplings \cite{VSECT}.  For the scalar 
leptoquark case, cross sections calculated at NLO with $\mu=2M_{\text{LQ}}$ 
are approximately equal to those calculated at LO with $Q^2 = M_{\text{LQ}}^2$. 
We therefore compare our cross section limit with LO calculations of vector 
leptoquark cross sections for this choice of $Q^2$ scale.

\begin{figure}
\includegraphics[width=3.25in]{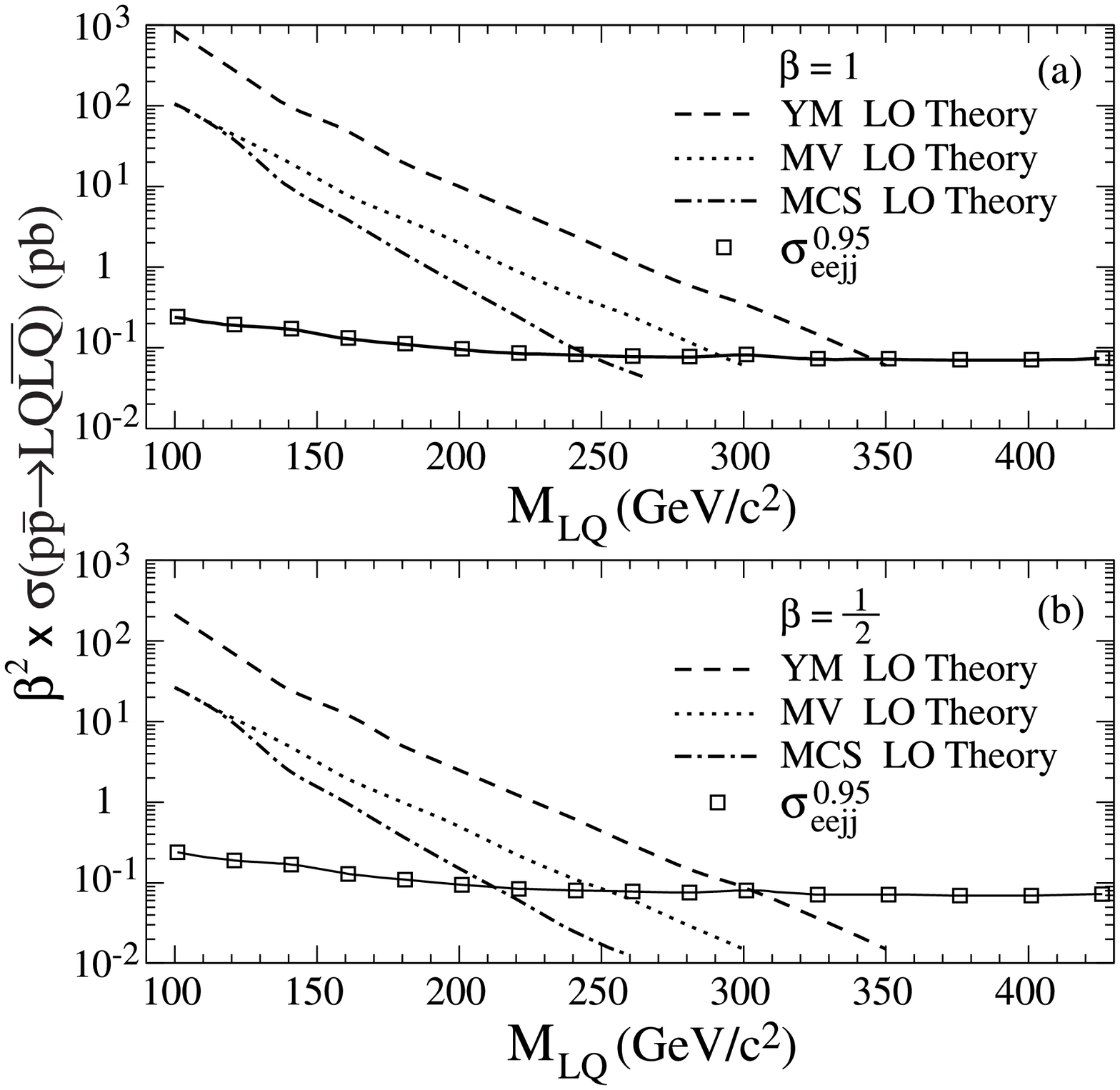}
\caption{The 95\% C.L. upper limits on the vector leptoquark pair 
production cross section from the $eejj$ channel and the LO predictions for 
Yang-Mills (YM), minimal vector (MV), and minimum cross section (MCS) 
couplings as a function of leptoquark mass for (a) $\beta = 1$ and 
(b) $\beta = \half$.}
\label{fig:vlqlimit_ee}
\end{figure}

Figure~\ref{fig:vlqlimit_ee}(a) shows the experimental limits along with the 
three theoretical vector leptoquark cross sections for the $eejj$ channel
for $\beta = 1$.  Here,
the experimental result yields a lower limit of $M_{\text{LQ}} > 340$~\gevcc\ 
for the vector leptoquarks assuming Yang-Mills coupling, 
$M_{\text{LQ}} > 290$~\gevcc\ for minimal vector coupling, and 
$M_{\text{LQ}} > 245$~\gevcc\ for the coupling corresponding to
the minimum cross section.
Similarly, for $\beta = \half$ (Fig.~\ref{fig:vlqlimit_ee}(b)), our 
result provides a lower limit of  
300~\gevcc\ for Yang-Mills coupling, 250~\gevcc\ for minimal vector coupling, 
and 210~\gevcc\ for the coupling corresponding to the minimum cross 
section.

\section{The {\lowercase{\textit{e}}}${\mathbf\nu}${\lowercase{\textit{jj}}} 
Channel}

For $0 < \beta < 1$, leptoquark pairs can decay to $e\nu jj$ as well as to 
$eejj$.  The $e\nu jj$ channel therefore allows us to 
extend the leptoquark mass limit to higher masses for $0 < \beta < 1$.  
Our optimization techniques for this analysis are similar to those
we used for the $eejj$ channel.

As in the $eejj$ channel, we use the same data sample for both the scalar  and 
vector-leptoquark analyses.  However, because the scalar-leptoquark analysis 
depends on a mass-based variable, and the vector leptoquark analysis is
sensitive to higher masses than the scalar leptoquark analysis, the final 
event selection is slightly different.  The scalar leptoquark analysis is 
described first, followed by the vector leptoquark analysis.

\subsection{The Data}

\subsubsection{Triggers}

The data sample for this analysis corresponds to an integrated luminosity 
of $115 \pm 6$~pb$^{-1}$.  Using events collected with the triggers shown in 
Table~\ref{table:evtrig}, the initial data sample contains 95,383 events. 

\begin{table*}
\setlength\tabcolsep{4pt}
\caption{The Level 2 triggers used in the $e\nu jj$ analysis.  The transverse
energy of an EM cluster is denoted by $E_T^{\text{EM}}$.  The number of
events is that in the initial data set.} 
\label{table:evtrig}
\begin{ruledtabular}
\begin{tabular}{@{}clcc}
Run &  Trigger Requirements & Integrated Luminosity & Number of Events\\
\hline
Run 1A & $E_T^{\text{EM}} > 20$ GeV & 11.2~pb$^{-1}$ &9,862 \\
Run 1B & $E_T^{\text{EM}} > 20$ GeV, isolated & 92.9~pb$^{-1}$ & 77,912\\
       & $\met > 15$~GeV \\
Run 1C &  $E_T^{\text{EM}} > 20$ GeV, isolated & 0.8~pb$^{-1}$ & 369\\
       & $\met > 15$~GeV \\
Run 1C &  $E_T^{\text{EM}} > 17$ GeV, isolated & 10.5~pb$^{-1}$ & 7,240\\
       & $E_T^{j_1,j_2} > 10$~GeV,\ $\met > 14$~GeV
\end{tabular}
\end{ruledtabular}
\end{table*}

\subsubsection{Event Selection for the Base Data Sample}

We require one electron with a matching track with $E_T^e > 30$~GeV, 
$\met > 20$~GeV, and at least two jets with $E_T^j > 20$~GeV.
Electrons with $E_T^e > 20$~GeV close to a jet 
($\Delta{\cal{R}}_e < 0.6$), are 
``subtracted'' from the jet in order not to double count the energy in the 
event.  Since the \met\ threshold for this analysis is relatively 
high, we use a ``minimal'' Main Ring veto to 
increase the efficiency (see Sec.~\ref{sec:MMRV}).  

\begin{figure}[ht] 
\includegraphics[width=2.35in]{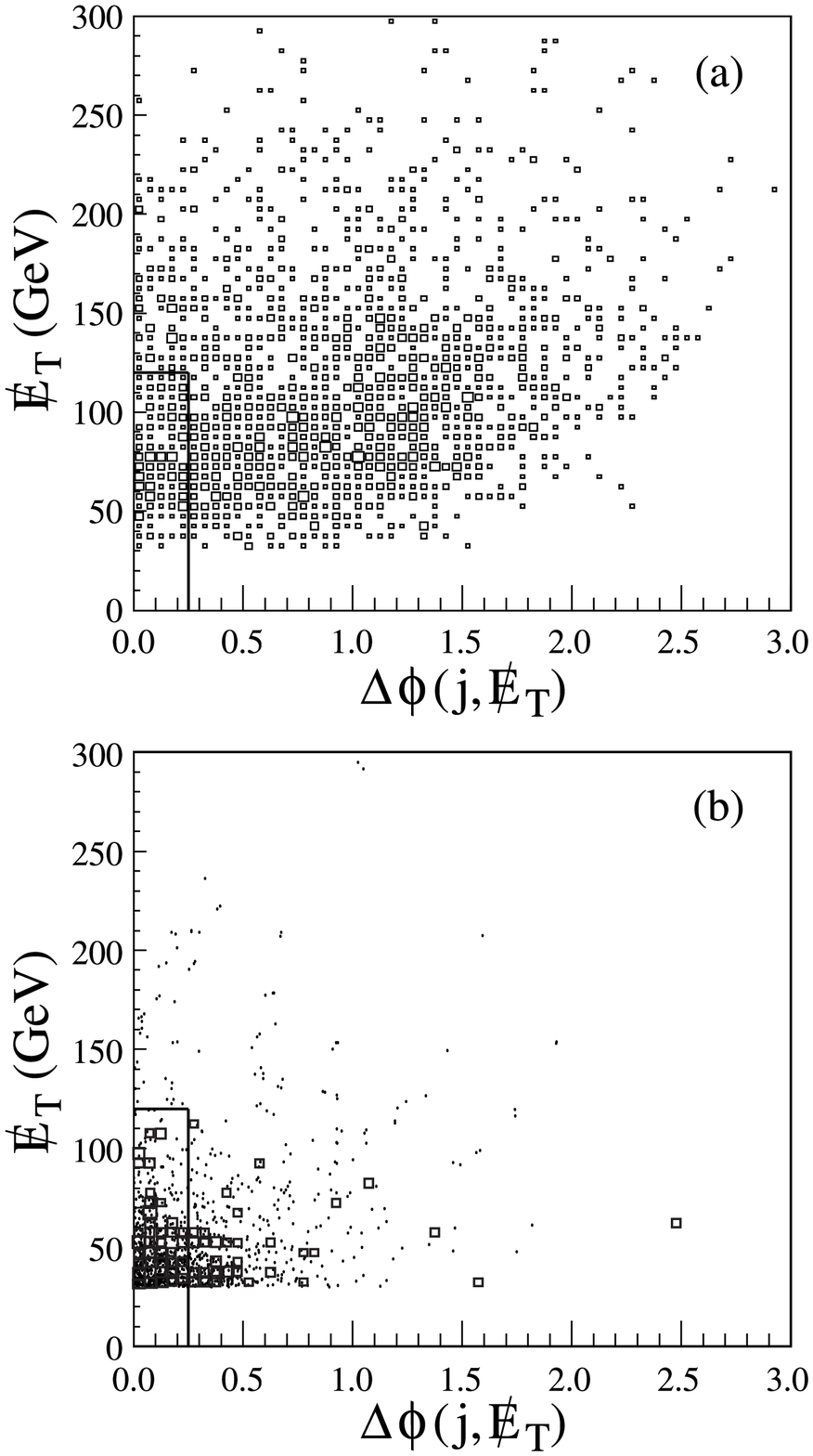}
\caption{Effect of the requirement of acolinearity in \met\ on (a) a 
180~\gevcc\ MC leptoquark signal and (b) the multijet background.  In (b), 
the dots show the distribution before imposition of the $M_T^{e\nu}$ 
requirement; the open squares show the distribution after applying the 
$M_T^{e\nu}$ requirement.  The acolinearity requirement is indicated by the 
solid lines.}
\label{fig:met-iso}
\end{figure}

To suppress the background from top-quark pair production, we apply a muon veto 
by requiring events to contain no well-reconstructed 
muons with $p_T > 4$~GeV/$c$  \cite{d0_topsearch_prd}.
To reduce the multijet background when $\met < 120$~GeV, we require the 
\met\ vector to be isolated in $\phi$ from any jets 
($\Delta\phi(j,\met) > 0.25$).  The effect of this requirement on a
180~\gevcc\ leptoquark MC sample and on the multijet background is shown in
Fig.~\ref{fig:met-iso}.

After the above cuts, 1094 events remain in the data sample, primarily from
$W+2j$ production.  To remove these events, we require 
$M_T^{e\nu} > 110$~\gevcc, reducing our base data sample to 14 
events. 

%limit\clearpage\newpage

\subsection{MC Signal Samples}

We use the {\footnotesize ISAJET} event generator followed by the full 
detector simulation via {\footnotesize GEANT} to model the leptoquark
signal.  Two thousand to five thousand events were generated in steps of
20~\gevcc\ for $M_{\text{LQ}}$ between 80 and 220~\gevcc.  We also use a
{\footnotesize PYTHIA}  MC sample at 200~\gevcc\ for studying MC systematics 
and for cross checks.

\subsection{Background Samples}
\label{sec:enujj_background}

As implied above, the dominant background to the $e\nu jj$ final state is 
$W+2j$ production.  The other significant backgrounds are from \ttbar\ 
production and multijet events in which a jet is misidentified as an 
electron and the energy is mismeasured, thereby introducing false \met.

\subsubsection{\ttbar\ Background}

The \ttbar\ MC event sample contains all leptonic final states for 
$m_t = 170$~\gevcc.  It was generated using {\footnotesize HERWIG} 
followed by {\footnotesize GEANT} detector simulation.  The sample 
corresponds to an integrated luminosity of about 32~fb$^{-1}$.

Since top-quark events frequently contain muons from $W \to \mu\nu$
and $b$-quark decays, the muon-veto requirement provides an effective way to
remove \ttbar\ events.  To determine the background due to top quark events, 
we apply all of the basic cuts except the muon and minimal Main Ring vetoes to 
the MC sample.  

Because the reconstruction efficiency for muons in MC events is higher than 
that for real muons, {\footnotesize GEANT} overestimates the rejection factor 
against muons.  The correction (between 50\% and 90\%) to the efficiency 
depends on the run number (due to chamber aging and repair) and the 
pseudorapidity of the muon.  
After applying this factor and the efficiencies described below, we estimate
that the data sample of 1094 events (before imposing the $M_T^{e\nu}$ cut) 
contains $12 \pm 4$ \ttbar\ events.  After requiring $M_T^{e\nu} > 110$~\gevcc, 
$2.0\pm 0.7$ \ttbar\ events are expected to remain in the base data sample of 
14 events.

\subsubsection{Multijet Background}

The multijet background is estimated using the data samples and the 
misidentification probability of $(3.50 \pm 0.35)\times 10^{-4}$ described in
Sec.~\ref{sec:QCD_misid}.  We select events from the multijet data sample that 
have at least three jets and $\met > 30$~GeV.  To minimize luminosity 
dependence and the misidentification of primary interaction vertices, we use 
only those events that have a single interaction vertex within the fiducial 
region of the detector ($|z_{\text{VTX}}| \le 50$\ cm).  To account for 
multiple interactions and multiple vertices, we apply a correction factor.  
The correction factor is determined by measuring the fraction of 
single-interaction events in the $Z+2j$ data sample as a function of 
luminosity, and then weighting this fraction with a luminosity profile of the 
multijet data stream.  The correction factor is $2.2\pm 0.2$. 

We next examine all three-jet combinations for each event.  We treat each
jet as an electron in turn and require each permutation to pass our electron 
and jet kinematic and fiducial requirements.  Since the misidentification rate 
already accounts for the probability for a jet to be misidentified as an
electron, we do not apply the electron identification criteria here.  The 
multijet background is then defined by the product of the number of
combinations that pass all criteria, the misidentification probability, and a 
factor that scales the multijet sample luminosity to the luminosity of the 
data.  There are $75 \pm 15$ events expected in the sample of 1094 events 
before the $M_T^{e\nu}$ cut and $4.1 \pm 0.9$ multijet events after the 
$M_T^{e\nu} > 110$~\gevcc\ requirement.  The uncertainty in the background 
accounts for the statistics of the multijet sample and for a 20\% systematic 
error reflecting the variation of the 
misidentification probability with $E_T$ and pseudorapidity, as well as jet 
trigger turn-on effects and the uncertainty in the scaling factor.

\subsubsection{$W+2j$ Background}

For the $W+2j$ background, we use a sample of events generated with
{\footnotesize VECBOS} \cite{VECBOS} followed by {\footnotesize ISAJET} 
underlying-event modeling and {\footnotesize GEANT} detector simulation.  This 
initial sample contains 227,726 events and corresponds to an integrated 
luminosity of approximately 0.8~fb$^{-1}$.

For calculating the background, the number of MC $W+2j$ events with 
$M_T^{e\nu} < 110$~\gevcc\ is normalized to the observed number of events after 
subtracting the estimated \ttbar\ and multijet backgrounds.    
A scaling factor of $0.22\pm 0.01$ gives good agreement between the 
Monte Carlo and the data and is consistent with the value of 0.20 expected 
from cross section and efficiency calculations. 

To check the normalization, we repeat
the comparison between the estimated background and the data for two additional
thresholds on the \met: $\met > 25$~GeV and $\met > 35$~GeV.  The agreement is 
again very good, showing that the fractional backgrounds are well-understood 
(the multijet background varies by a factor of six, from 115 to 20 events, 
between the two thresholds). 

The number of $W+2j$ events in the base data sample is estimated to be 
$11.7 \pm 1.8$ events.

\subsubsection{Total Background}

\begin{figure} 
\includegraphics[width=3.25in]{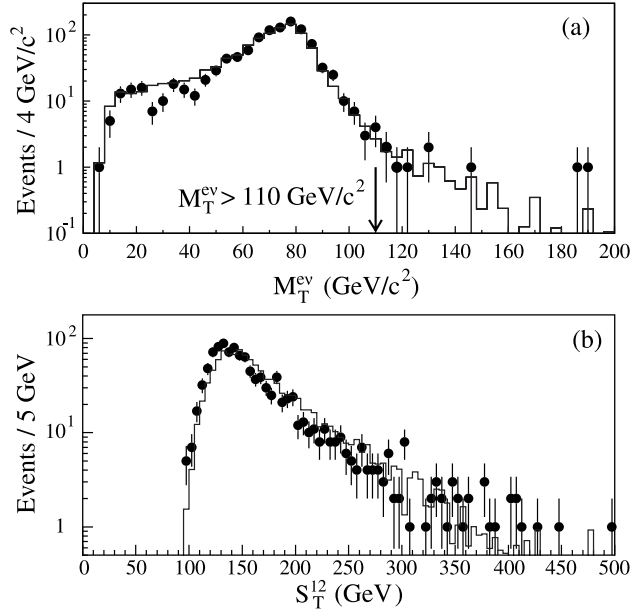}
\caption{Comparison of the (a) $M_T^{e\nu}$ and (b) $S_T^{12}$ distributions 
for the $e\nu jj$ data (points with error bars) and the predicted background 
(solid histogram) before imposing the $M_T^{e\nu}$ requirement.}
\label{fig:mt_st12}
\end{figure}

Figures~\ref{fig:mt_st12}(a) and \ref{fig:mt_st12}(b) show the 
$M_T^{e\nu}$ and $S_T^{12}$ distributions for the data sample and the 
background before the cut on $M_T^{e\nu}$.
It is clear that we model the transverse mass distribution quite well up to 
110~\gevcc.  The $S_T^{12}$ distribution is also well-described by 
the MC except for the small systematic offset of the
prediction relative to the data.

The total background estimate after basic requirements is $17.8 \pm 2.1$ 
events, in agreement with the 14 events observed in the data.

\subsection{Efficiencies}

\subsubsection{Trigger Efficiency}

Since events in the base data sample are required to have high electron \et\ 
and \met, the trigger requirements listed in Table~\ref{table:evtrig} are 
very 
efficient.  The EM part of the trigger has an efficiency of $(99.5 \pm 0.5)$\%. 

\subsubsection{Efficiency of the Minimal Main Ring Veto}
\label{sec:MMRV}

As discussed in Sec.~\ref{sec:detector}, additional Main Ring (MR) trigger 
requirements can be applied offline to events collected using triggers with 
liberal MR requirements.  For the $e\nu jj$ analysis, we apply a ``minimal'' MR 
veto to remove events that occurred during proton injection and when 
the proton bunch passed through the detector, while keeping events collected 
during the calorimeter recovery period.  The efficiency of this veto is 
estimated using $Z + 2j$ data collected using triggers with looser MR 
requirements than in the triggers used in the $e\nu jj$ analysis.  First, the 
MR requirements for the $e\nu jj$ triggers are applied to the $Z + 2j$ data.  
The efficiency of the minimal MR veto is then calculated by comparing the 
number of events in the $Z$-boson mass peak before and after the additional 
minimal MR veto requirements are applied.  The efficiency of this veto is 
($94 \pm 1$)\% (i.e. 6\% of the good events are removed along with a much 
larger percentage of background events).  If the ``calorimeter recovery'' 
events were also removed, the efficiency would be reduced to about 90\%.  

\subsubsection{Muon-Veto Efficiency}

The efficiency of the muon veto is estimated using a sample of 
$Z(\to ee) + 2j$ events. 
Except for the additional electron, these events have a topology similar to that
of leptoquark events in the $e\nu jj$ channel and should have a similar random 
muon track rate.  The calculation is
done using the number of events in the $Z$-boson mass peak before and after
application of the muon veto.  Background under the $Z$ boson is subtracted
using the standard side-band technique.  The muon-veto efficiency is 
($97 \pm 1)$\%.

\subsubsection{Electron Identification Efficiencies}

Using the efficiencies described in Sec.~\ref{sec:e_eff} for the $eejj$ 
channel, the overall electron
identification efficiency for leptoquark events in the $e\nu jj$ channel is
($61 \pm 4$)\% in the CC and ($54 \pm 4$)\% in the EC.  Since 
($93 \pm 1$)\% of the electrons in the $e\nu jj$ final state are in the 
CC, the total electron identification efficiency, including tracking and 
quality requirements, is $(60 \pm 3)$\%.

\subsection{Event Selection Optimization}
\label{sec:enjj_opt}

\subsubsection{Random Grid Search}

We use a random grid search based on the $M_{\text{LQ}} = 180$~\gevcc\ MC 
sample to select the optimal variables and thresholds for the $e\nu jj$ 
channel.  Many different variables and combinations of variables (see
Sec.~\ref{sec:variables}) were tested for their 
efficiency in retaining the signal and rejecting the background.  The inputs 
to the {\footnotesize RGSEARCH} program are the MC signal samples 
and the $W+2j$, \ttbar, and multijet background samples described in
Sec.~\ref{sec:enujj_background}.  The combinations of variables that have 
the most discriminating power are then used in the neural network analysis.  
The most powerful variables for separating leptoquark signals from the 
background are $S_T^{12}$ and $\frac{\delta M}{M}(M_{\text{LQ}})$ 
(see Sec.~\ref{sec:variables}).

\subsubsection{Neural Network Analysis}

\begin{figure}
\includegraphics[width=3.35in]{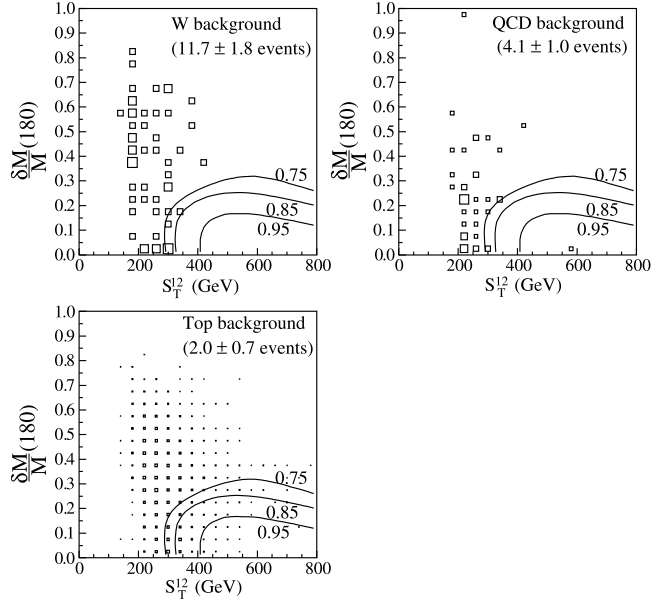}
\caption{Distributions of $\frac{\delta M}{M}(180)$ versus $S_T^{12}$ for the 
three individual backgrounds: (a) $W+2j$ events, (b) multijet events, and 
(c) \ttbar\ events.   The curves show neural net contours for 
${\cal D}_{\text{NN}}(180) = 0.75, 0.85,\ {\text{and}}\ 0.95$.}
\label{fig:nn-3bkg}
\end{figure}

\begin{figure}
\includegraphics[width=3.35in]{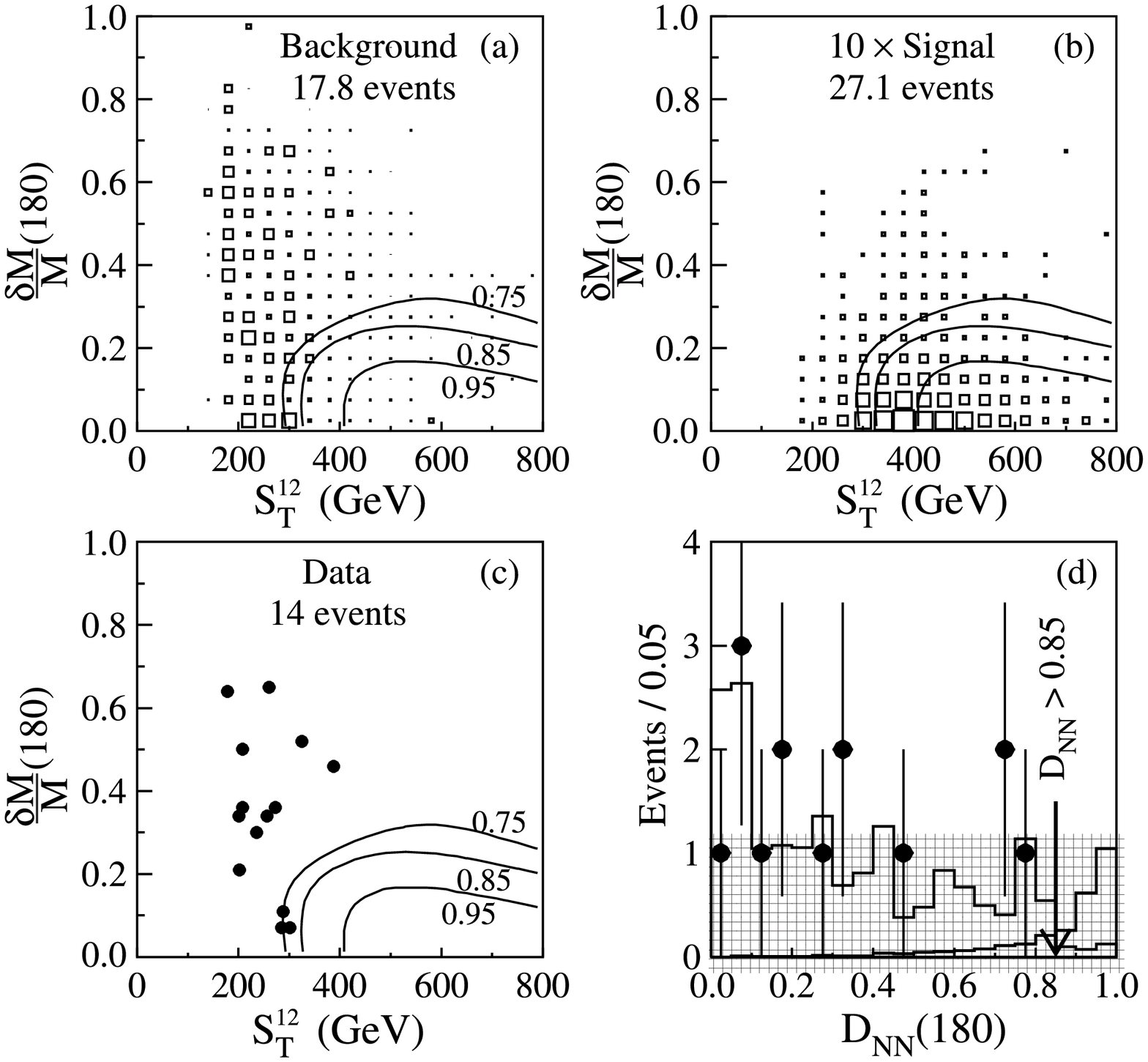}
\caption{Distributions of $\frac{\delta M}{M}(180)$ versus $S_T^{12}$ for 
(a) the total background, (b) ten times the expected signal from 180~\gevcc\ 
leptoquarks, and (c) the data. (d) The neural network discriminant for the 
signal (hatched histogram), the background (open histogram), and the data 
(points with error bars).  The curves show neural net contours for 
${\cal D}_{\text{NN}} = 0.95, 0.85,\ {\text{and}}\ 0.75$.}
\label{fig:nn-res}
\end{figure}

We use a neural network with two input nodes (corresponding to the 
variables $S_T^{12}$ and $\frac{\delta M}{M}(M_{\text{LQ}})$), five hidden 
nodes, and one output node.  A separate network is trained for each MC signal 
sample (with a desired network output 
${\cal D}_{\text{NN}}(M_{\text{LQ}}) = 1$) and 
the expected admixture of $W+2j$, \ttbar, and multijet background 
events (with desired ${\cal D}_{\text{NN}}(M_{\text{LQ}}) = 0$).
The expected rejection can be seen in Figs.~\ref{fig:nn-3bkg} and 
\ref{fig:nn-res}.  Figure~\ref{fig:nn-3bkg} shows the two-dimensional 
distributions of
$\frac{\delta M}{M}(180)$ versus $S_T^{12}$ for the three individual 
backgrounds.  Figures~\ref{fig:nn-res}(a)--(c) show the same two-dimensional 
distributions for the total background, simulated leptoquark events with 
$M_{\text{LQ}} = 180$~\gevcc, and the data.  The contours corresponding to 
constant values of ${\cal D}_{\text{NN}}(180) = 0.75$, 0.85, and 0.95 
demonstrate the level of separation achieved between the expected signal and 
the background.  The distribution of
${\cal D}_{\text{NN}}(180)$ for the data is compared with the predicted
distributions for background and signal in Fig.~\ref{fig:nn-res}(d).  The data 
can be described by background alone.  The highest
value of ${\cal D}_{\text{NN}}(180)$ observed in the base data sample is 0.79.  

Using the strategy described in Sec.~\ref{sec:optimization_criterion}, 
we optimize the signal for a fixed background of approximately 0.4 events.  
In the low-mass range ($M_{\text{LQ}} \le 120$~\gevcc), where leptoquark 
production rates are high, requiring $S_T^{12} > 400$~GeV is sufficient and
leads to a background of $0.60 \pm 0.27$ events, consistent with the desired
background level.  For
$M_{\text{LQ}} > 120$~\gevcc, we use neural networks since they provide
higher efficiency than an $S_T^{12}$ cut alone.  For 180~\gevcc\
leptoquarks, approximately 0.4 background events are expected for 
${\cal D}_{\text{NN}}(180) > 0.85$.  We choose the 
${\cal D}_{\text{NN}}(M_{\text{LQ}})$ threshold to be a multiple of 0.05 
rather than a value that yields exactly 0.4 background events; 
${\cal D}_{\text{NN}}(180) > 0.85$ corresponds to a background of 
$0.29 \pm 0.25$ events.  No events in the base data sample satisfy this
criterion.  Naturally, for leptoquark masses other than 180~\gevcc, the
requirement on ${\cal D}_{\text{NN}}(M_{\text{LQ}})$ is different.  The 
expected background varies between 0.29 and 0.61 events and is listed in
Table~\ref{table:cs-ev}.  No events from the 
base data sample pass any of these ${\cal D}_{\text{NN}}(M_{\text{LQ}})$
thresholds.

\begin{table}
\setlength\tabcolsep{4pt}
\caption{Efficiency, background, 95\% C.L. upper limit on the leptoquark 
production cross section, and NLO cross section multiplied by the branching
fraction with $\mu=2M_{\text{LQ}}$~\protect{\cite{kraemer}} for $\beta=\half$ 
as a function of leptoquark mass for the $e\nu jj$ channel.}
\label{table:cs-ev}
\begin{ruledtabular}
\begin{tabular}{@{}ccccc}
{\text{Mass}} & Efficiency & Background & $\sigma_{\text{limit}}$ & 
                             $2\beta(1-\beta) \sigma_{\text{NLO}}$ \\
(\gevcc) & (\%) & (Events) & (pb) & (pb) \\ 
\hline
80 &  $0.32 \pm 0.08$ & $0.60 \pm 0.27$ & 10.9 & 18.0\\
100 & $1.15 \pm 0.21$ & $0.60 \pm 0.27$ & 2.6 & 5.34 \\
120 & $2.45 \pm 0.33$ & $0.60 \pm 0.27$ & 1.0 & 1.90 \\
140 & $6.65 \pm 0.96$ & $0.54 \pm 0.25$ & 0.43 & 0.77\\
160 & $10.9 \pm 1.2 $ & $0.61 \pm 0.27$ & 0.24 & 0.34\\
180 & $14.7 \pm 1.2 $ & $0.29 \pm 0.25$ & 0.18 & 0.16\\
200 & $19.4 \pm 1.7 $ & $0.43 \pm 0.27$ & 0.14 & 0.08\\
220 & $21.5 \pm 1.7 $ & $0.41 \pm 0.27$ & 0.12 & 0.04
\end{tabular}
\end{ruledtabular}
\end{table}

Rectangular cuts of $S_T^{12} > 350$~GeV and $\frac{\delta M}{M}(180) < 0.25$ 
yield a total background of 0.4 events.  This also leaves no events 
in the data sample, but the signal efficiency is approximately 10\% lower for
$M_{\text{LQ}} = 180$~\gevcc.

\subsection{Check}

\begin{figure}
\includegraphics[width=3.25in]{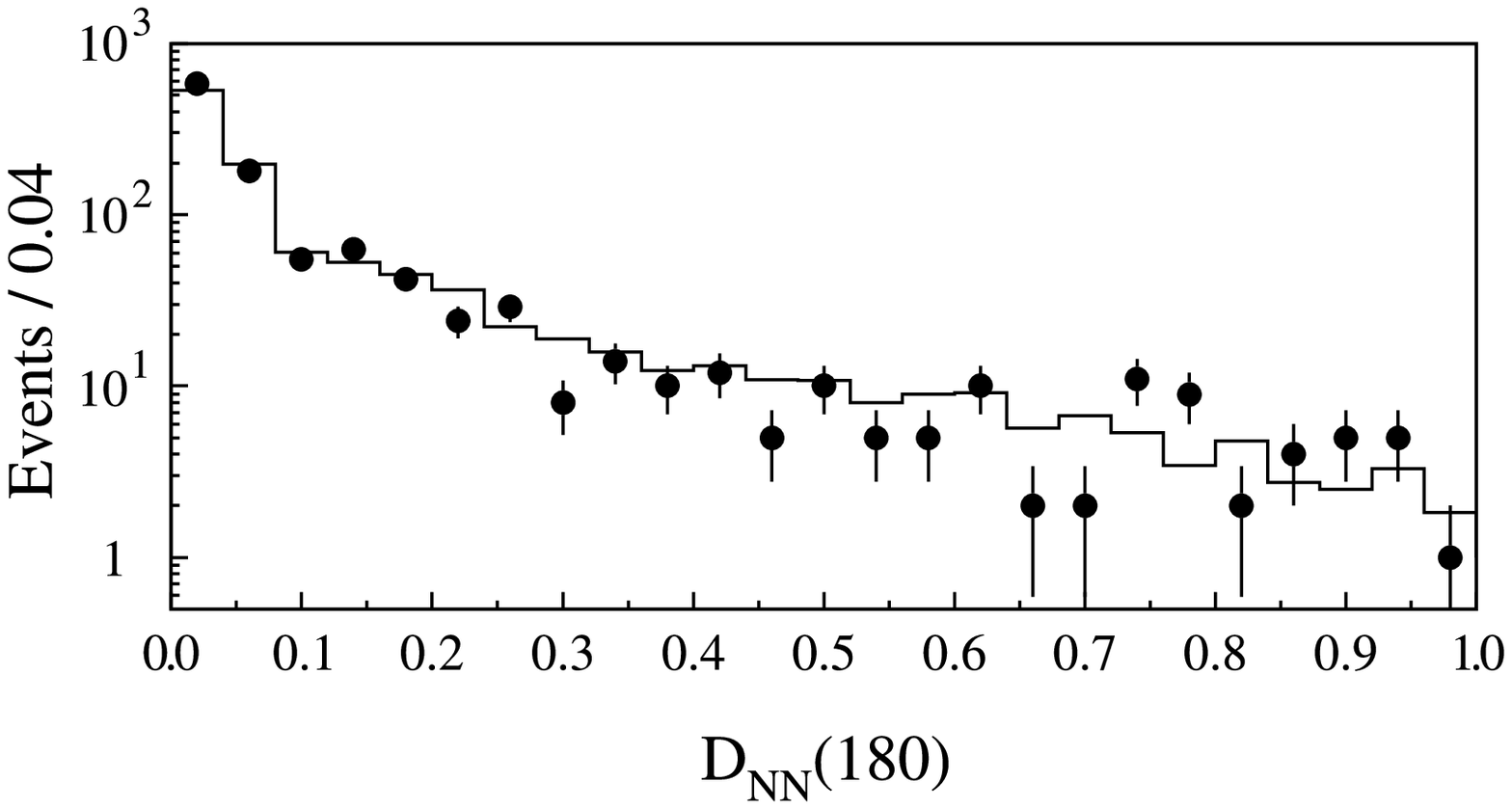}
\caption{Comparison of the ${\cal{D}}_{\text{NN}}(180)$ distribution for the 
$e\nu jj$ data (points with error bars) and the predicted background 
(solid histogram) before the cut on $M_T^{e\nu}$.}
\label{fig:enujj_dnn}
\end{figure}

As a check of our understanding of the background, Fig.~\ref{fig:enujj_dnn} 
shows the distribution of ${\cal{D}}_{\text{NN}}(180)$ for the data 
and for the predicted background 
before the $M_T^{e\nu}$ cut.  The agreement is acceptable.

\subsection{Signal Studies}

\subsubsection{Systematic Uncertainties}
 
The systematic uncertainty in the signal efficiency varies from 25\% to 8\% 
for $M_{\text{LQ}}$ between 80 and 220~\gevcc.  The sources and sizes of the
systematic uncertainties are given in 
Table~\ref{table:evsyst}.  The uncertainties due to the jet energy scale 
and initial/final state radiation are significantly lower than in the $eejj$ 
analysis due to the use of $S_T^{12}$ rather than the all-jets-based $S_T$ as a
discriminator.

\begin{table}
\caption{Systematic uncertainties in the signal for the $e\nu jj$ analysis.}
\label{table:evsyst}
\begin{ruledtabular}
\begin{tabular}{@{}ll}
Source of Systematics & Uncertainty (\%)\\
\hline
Particle identification & 5 \\
Smearing in the detector & 3 \\
Jet energy scale & 10--2 ($M_{\text{LQ}} = 80$--220~\gevcc) \\
Gluon radiation & 4 \\
PDF and $Q^2$ scale & 5 \\
Monte Carlo statistics & 25--3 ($M_{\text{LQ}} = 80$--220~\gevcc)\\
Luminosity & 5 \\
\hline
Total & 25--8 ($M_{\text{LQ}} = 80$--220~\gevcc) \\
\end{tabular}
\end{ruledtabular}
\end{table}

\subsubsection{Signal Efficiency}

The signal detection efficiencies are calculated using simulated leptoquark 
events that pass the selection requirements; they
are shown in Table~\ref{table:cs-ev}.  The errors in the signal 
efficiencies include uncertainties in trigger and particle-identification 
efficiencies, the jet energy scale, effects of gluon radiation and parton 
fragmentation in the signal modeling, and finite MC statistics.

\subsection{Results from the {\mbox{$e\nu jj$}} Channel for Scalar Leptoquarks}

\begin{figure}
\includegraphics[width=3.0in]{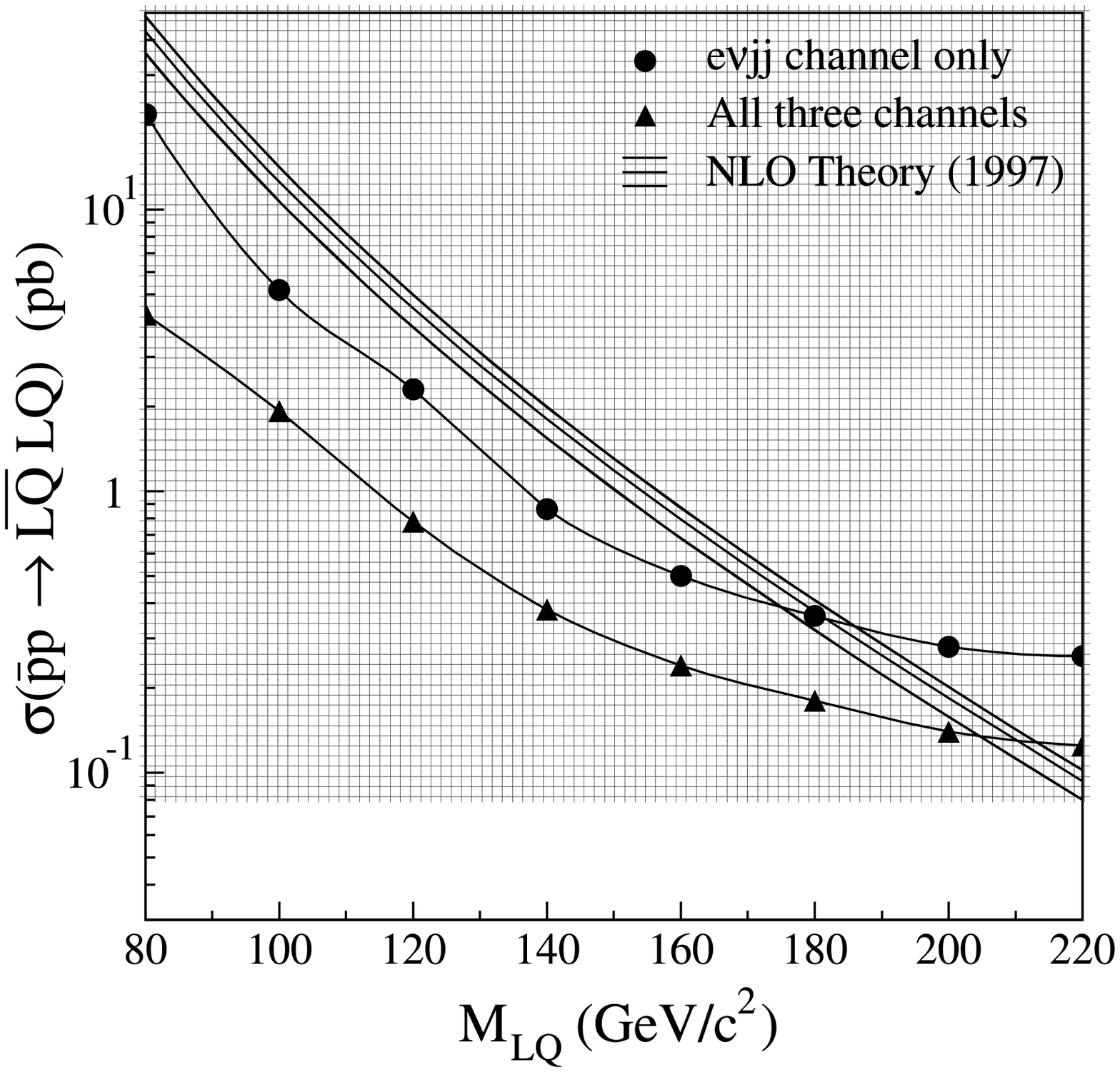}
\caption{The 95\% C.L. upper limits on the cross sections for scalar leptoquark 
pair production from the $e\nu jj$ channel, and for all three channels 
combined, for $\beta = \half$, compared to the NLO prediction, as a function of
leptoquark mass.}
\label{fig:slq_betahalf_lim}
\end{figure}

We obtain a 95\% C.L. upper limit on the scalar
leptoquark pair-production cross section for $\beta= {1 \over 2}$ as a
function of leptoquark mass.  The results, based on a Bayesian analysis 
\cite{limit_setting}, are shown in
Table~\ref{table:cs-ev}.  The statistical and systematic uncertainties in the
efficiency, the integrated luminosity, and the background estimation are
included in the limit calculation, all with Gaussian priors.  The
95\% C.L. upper limits on the cross section for scalar leptoquark pair 
production in the $e\nu jj$ channel,
corrected for the branching fraction of $\beta= {1 \over 2}$, 
for various leptoquark masses are plotted in Fig.~\ref{fig:slq_betahalf_lim} 
along with the NLO calculations \cite{kraemer}.  The intersection of our 
limit with the lower edge of the theory band (renormalization
scale $\mu = 2M_{\text{\text{LQ}}}$) is at 0.38~pb, leading to a 95\% C.L.
lower limit on the leptoquark mass of 175~\gevcc.

\subsection{Vector Leptoquarks}

As in the case of the $eejj$ channel, vector leptoquark events were 
generated for $M_{\text{LQ}}$ between 100 \gevcc\ and 425 \gevcc\ using a 
version of {\footnotesize PYTHIA} modified to include vector leptoquarks with 
different couplings.
% \cite{msu}.  
The distributions of the kinematic variables for 
scalar and vector leptoquarks are similar, and consequently, the same 
event selection is used for both analyses for $M_{\text{LQ}} \leq 220$~\gevcc.

Neural networks for the $e\nu jj$ channel were trained on scalar leptoquark 
MC samples up to $M_{\text{LQ}} = 220$~\gevcc.  Since vector leptoquark 
production cross sections are higher than scalar leptoquark cross sections, 
higher masses are of more interest.  For vector leptoquarks with 
$M_{\text{LQ}} > 220$~\gevcc, we require 
$S_T^{12} > 400$~GeV.  This variable is one of the inputs to the neural network 
and provides good signal identification efficiency and a background of 
$0.60 \pm 0.27$ events.

\begin{figure}
\includegraphics[width=3.0in]{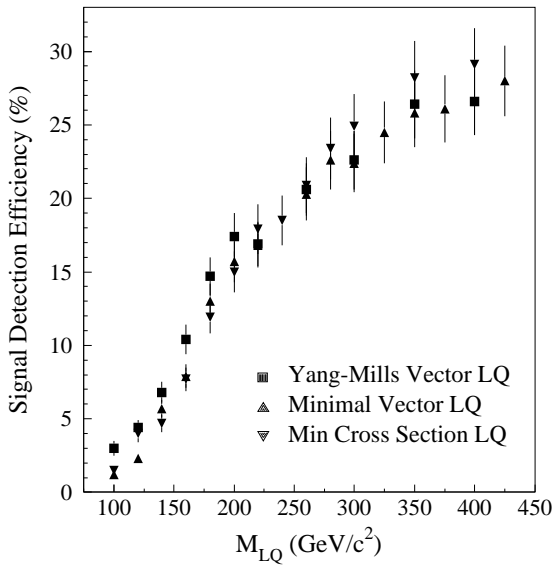}
\caption{The detection efficiency for vector leptoquarks in the $e\nu jj$ 
channel.}
\label{fig:eveff}
\end{figure}

Again, the identification efficiencies for vector leptoquarks for the three 
couplings agree within their uncertainties, as shown in 
Fig.~\ref{fig:eveff}.  Therefore, to reduce the statistical uncertainty in our 
analysis, we use the average identification efficiency of the three sets of MC
events to set a single experimental limit on the cross section.  As before, 
this limit is compared with the appropriate theoretical cross section for each 
coupling.

\begin{figure}
\includegraphics[width=3.0in]{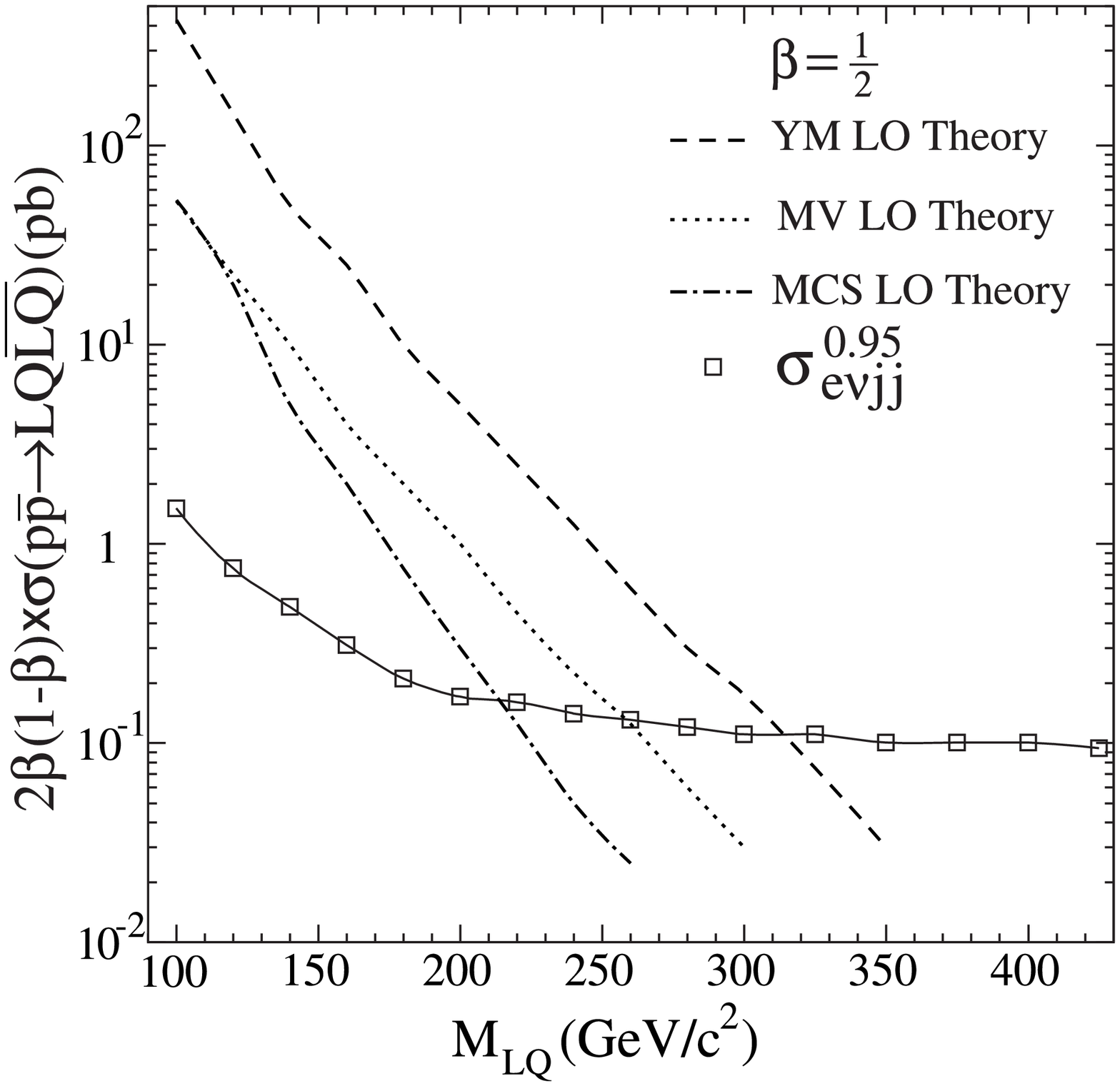}
\caption{The 95\% C.L. upper limits on the cross sections for vector leptoquark 
pair production from the $e\nu jj$ channel for $\beta = \half$, and the LO  
predictions for the three couplings, as a function of leptoquark mass.}
\label{fig:vlqlimit_05ev}
\end{figure}

Figure~\ref{fig:vlqlimit_05ev} shows the experimental limits along with the 
three theoretical LO vector leptoquark cross sections \cite{VSECT} for the 
$e\nu jj$ channel for $\beta = \half$ and $Q^2 = M_{\text{LQ}}^2$.  
For Yang-Mills coupling, the experimental lower limit on the vector leptoquark 
mass is 315~\gevcc, for $\beta = \half$.  For minimal 
vector coupling, the mass limit is 260~\gevcc\ for $\beta = \half$.
For the coupling corresponding to the minimum
cross section, the lower limit is 215~\gevcc\ for $\beta = \half$.

\section{The {${\mathbf\nu\mathbf\nu}${\lowercase{\textit{jj}}} channel}}

To analyze the $\nu\nu jj$ channel, we make use of our published search 
\cite{stop} for the supersymmetric partner of the top quark using just the 
1992--1993 data sample.  In that analysis, we searched for the pair production 
of top squarks that decay exclusively via a $c$ quark and the lightest
neutralino, $\widetilde{t}_1 \to c\widetilde{\chi}_1^0$, resulting in a final 
state with \met\ and two acolinear jets.   

Approximately 75\% of the data were collected using a trigger whose 
primary  requirement was $\met > 35$~GeV at Level 2; the balance had a \met\ 
threshold of 40~GeV.
To ensure an unambiguous \met\ measurement, events were required to have only
one primary vertex, reducing the sample to single interactions with an
integrated luminosity equivalent to approximately $7.4$~pb$^{-1}$. 

Events were required to have $\met >40$~GeV, two jets with $E_T>30$~GeV, and 
no isolated electrons or muons with $E_T > 10$~GeV.  
In addition, the two leading jets were 
required to be acolinear ($90^\circ < \Delta\phi(j_1,j_2) <165^\circ $), and 
the \met\ was required not to be aligned with either the leading jet
($10^\circ < \Delta\phi(j_1,\met) <125^\circ$) or the third or fourth 
leading jets ($10^\circ < \Delta\phi(j_{3,4},\met)$).
Three events survived the selection criteria, consistent with
the estimated background of $3.5\pm 1.2$ events, primarily from $W$/$Z$+jets
production.  

The efficiencies of the event selection for scalar leptoquarks with 
$M_{\text{LQ}}$ from 50 to 200~\gevcc\ are calculated 
using signal MC events generated with the
{\footnotesize ISAJET} generator and processed through the
{\footnotesize GEANT}-based detector simulation.  The systematic errors in
the signal acceptance are calculated as in Ref. \cite{stop}.  The 
efficiencies, background, and cross section limits 
are shown in Table~\ref{table:cs-vv}.
This analysis yields the limit $M_{\text{LQ}}>79$~\gevcc\ at the 95\% C.L. for 
$\beta=0$.

\begin{table}
\setlength\tabcolsep{6pt}
\caption{Efficiency, background, 95\% C.L. upper limit on the leptoquark pair
production cross section, and the NLO cross section with 
$\mu=2M_{\text{LQ}}$~\protect{\cite{kraemer}} for $\beta=0$ as a function of 
leptoquark mass for the $\nu\nu jj$ channel.}
\label{table:cs-vv}
\begin{ruledtabular}
\begin{tabular}{@{}ccccc}
Mass & Efficiency & Background & $\sigma_{\text{limit}}$ & 
                                                      $\sigma_{\text{NLO}}$ \\ 
(\gevcc) & (\%) & (Events) & (pb) & (pb) \\ 
\hline
 50 & $0.446 ^{+0.096}_{-0.107}$ & $3.49 \pm 1.17$ & 328 & 406 \\
 60 & $1.11 \pm 0.16$          & $3.49 \pm 1.17$ & 77.0 & 162\\
 80 & $2.15  ^{+0.23}_{-0.22}$ & $3.49 \pm 1.17$ & 37.7 & 36.0\\
100 & $3.90 \pm 0.30$          & $3.49 \pm 1.17$ & 21.0 & 10.7\\
120 & $4.62 ^{+0.30}_{-0.32}$    & $3.49 \pm 1.17$ & 17.6 & 3.81\\
140 & $6.07 \pm 0.34$          & $3.49 \pm 1.17$ & 13.2 & 1.54\\
160 & $6.15 \pm 0.34$          & $3.49 \pm 1.17$ & 13.0 & 0.68\\
200 & $6.36 ^{+0.35}_{-0.36}$    & $3.49 \pm 1.17$ & 12.6 & 0.16\\
\end{tabular}
\end{ruledtabular}
\end{table}

The identification efficiency for vector leptoquark (generated using 
{\footnotesize PYTHIA}) and scalar leptoquark 
events with $M_{\text{LQ}} = 200$~\gevcc\ are identical, within errors.  Based
on this comparison, and similar comparisons in the $eejj$ and $e\nu jj$ 
channels, we use the experimental limit for scalar
leptoquarks for vector leptoquarks in the $\nu\nu jj$ channel.  Comparison
with the theoretical cross sections leads to 95\% C.L. limits of 
$M_{\text{LQ}}>$~206, 154, and 144~\gevcc\ for Yang-Mills, minimal vector, and
minimum cross section couplings, respectively, for $\beta = 0$.

\section{Gap in the Limit for Scalar Leptoquarks}

In our analysis of the $eejj$ and $e\nu jj$ channels, we use MC samples of 
leptoquarks with $M_{\text{LQ}} \ge 80$~\gevcc\ but our analysis is optimized
for leptoquarks with masses near 200~\gevcc.  From the $e\nu jj$ analysis, we
exclude $\beta > 0.13$ for $M_{\text{LQ}} = 80$~\gevcc.  The mass 
limit from the $\nu\nu jj$ channel for $\beta = 0.13$ is approximately 
75~\gevcc, leaving a small gap in our limit.  

To fill this gap, we examine further the 14 events in the
base data sample in the $e\nu jj$ analysis.  Making the very conservative 
assumption that all 14 events are due to leptoquark pair production, the 95\% 
C.L. upper limit on the cross section multiplied by the branching fraction and 
efficiency is 0.20~pb.  This permits us to extend our exclusion region to 
include $0.09 \le \beta \le 0.91$ for $M_{\text{LQ}} = 80$~\gevcc\ and 
$0.05 \le \beta \le 0.95$ for $M_{\text{LQ}} = 75$~\gevcc.  To obtain the
efficiency for $M_{\text{LQ}} = 75$~\gevcc, we scale the efficiency found for
higher $M_{\text{LQ}}$. 

\section{Combined Results}

\begin{figure}
\includegraphics[width=3.0in]{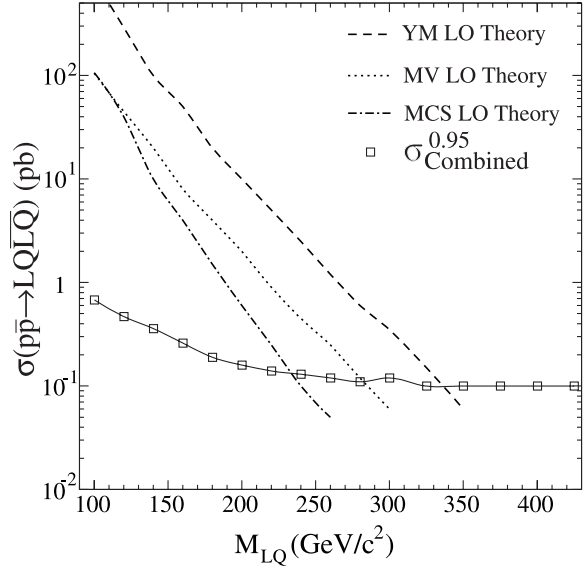}
\caption{The 95\% C.L. upper limits on cross sections for vector leptoquark 
pair production 
from all three channels combined for $\beta = \half$, and the LO 
predictions, as a function of leptoquark mass.}
\label{fig:vlim-comb}
\end{figure}

\begin{figure}
\includegraphics[width=3.25in]{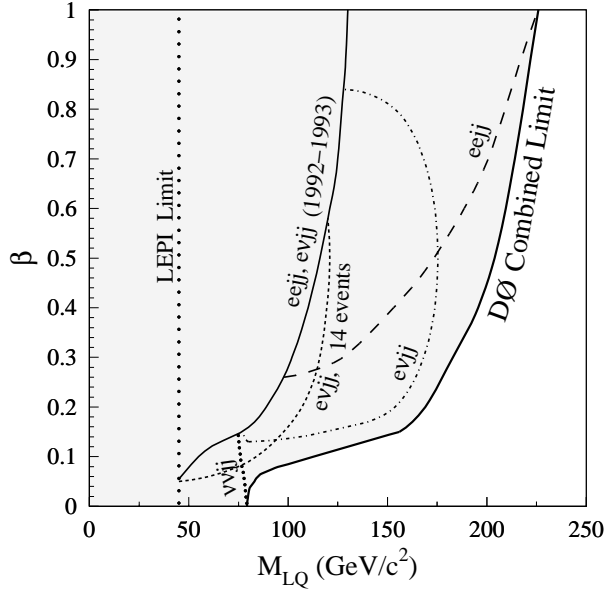}
\caption{The 95\% C.L. lower limit on the mass of first-generation scalar 
leptoquarks as a function of $\beta$ for the individual $eejj$, $e\nu jj$, and 
$\nu\nu jj$ channels, and for the combined analysis.}
\label{fig:mass_beta1}
\end{figure}

\begin{figure}
\includegraphics[width=3.15in]{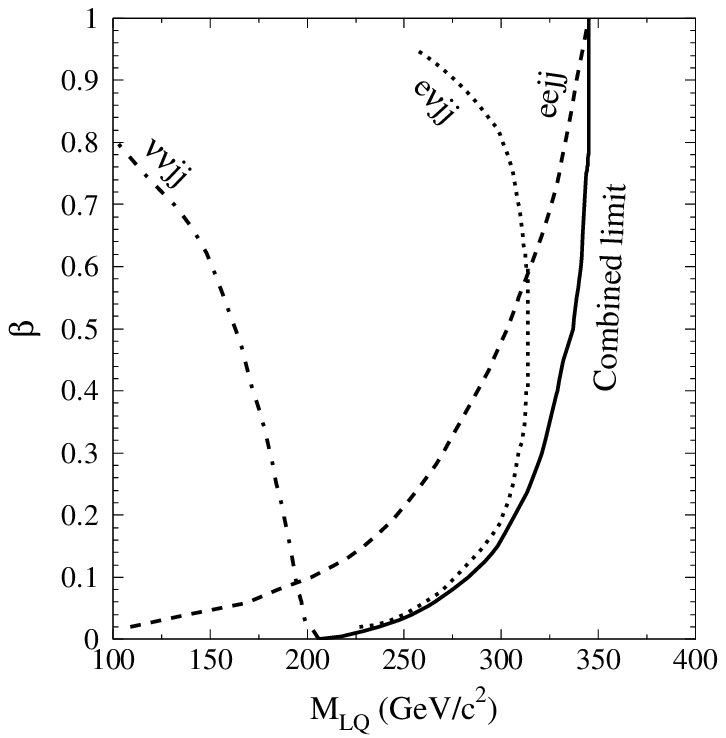}
\caption{The 95\% C.L. lower limit, as a function of $\beta$, on the mass of 
first-generation vector leptoquarks with Yang-Mills couplings from the
individual $eejj$, $e\nu jj$, and $\nu\nu jj$ channels and for the combined 
analysis.}
\label{fig:vlq_ym_excl}
\end{figure}

\begin{figure}
\includegraphics[width=3.0in]{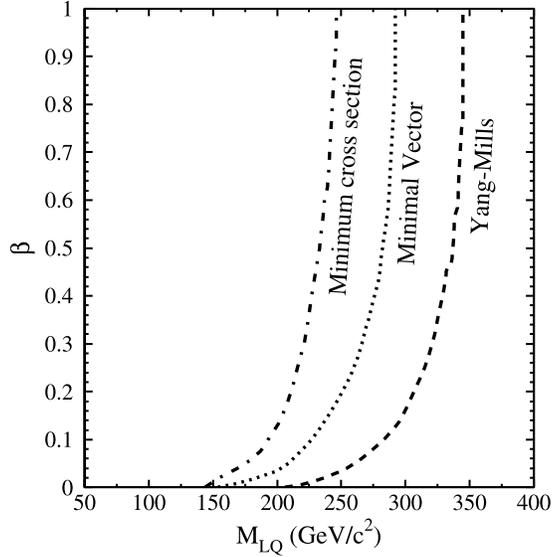}
\caption{The 95\% C.L. lower limits on $M_{\text{LQ}}$ as a function of $\beta$ 
for first-generation vector leptoquarks with Yang-Mills, minimal vector, and 
minimum cross section couplings from the $eejj$, 
$e\nu jj$, and $\nu\nu jj$ channels combined.}
\label{fig:vlq_exclusion}
\end{figure}

Combining \cite{limit_setting} the limits from the $eejj$, $e\nu jj$, and 
$\nu\nu jj$ channels, we obtain 95\% C.L. upper limits on the leptoquark 
pair-production cross section as a function of leptoquark mass and $\beta$.  
The cross-section limits for $\beta = \frac{1}{2}$ are shown in 
Fig.~\ref{fig:slq_betahalf_lim} for scalar leptoquarks
and in Fig.~\ref{fig:vlim-comb} for vector leptoquarks.
Table~\ref{table:mass_lims} lists the mass limits for 
$\beta=1,\ \frac{1}{2},\ {\text{and}}\ 0$
for the types of leptoquarks studied.  The lower limits on the mass of scalar
leptoquarks as a function of $\beta$, for all three channels combined, as 
well as for the individual channels, are shown in Fig.~\ref{fig:mass_beta1}.
Figure~\ref{fig:vlq_ym_excl} shows the exclusion contours 
from the individual channels and the combined result for vector leptoquarks 
with Yang-Mills coupling.  Figure~\ref{fig:vlq_exclusion} shows the overall 
exclusion contours for the three vector couplings.

\section{Conclusions}

We have presented 95\% C.L. upper limits on the pair production of
leptoquarks that decay to the $eejj$, $e\nu jj$ and $\nu\nu jj$ final states.
For scalar leptoquarks, the limits on the cross section provide  
lower limits on the scalar leptoquark mass of 225~\gevcc\ for $\beta = 1$, 
204~\gevcc\ for $\beta = \half$, and 79~\gevcc\ for $\beta = 0$.  
We have also set mass limits for vector leptoquarks for different couplings 
and have presented exclusion contours on 
$\beta$ and $M_{\text{LQ}}$.  At the 95\% C.L., our results exclude
an interpretation of the HERA high-$Q^2$ excess as $s$-channel scalar
leptoquark production for $M_{\text{LQ}} < 200$~\gevcc\ and $\beta > 0.4$.
These results can be also used to set limits on the pair production of any 
heavy scalar particle that decays into a lepton and a quark as expected in 
a variety of models and to restrict any new leptoquark models containing 
additional fermions
\cite{dont_stop}.

\begin{table*}
\setlength\tabcolsep{10pt}
\caption{Limits on the masses of first-generation leptoquarks.}
\label{table:mass_lims}
\begin{ruledtabular}
\begin{tabular}{@{}rrrrr}
$\beta$ & Scalar & Minimum Cross Section & Minimal Vector & Yang-Mills \\
 & (\gevcc) & (\gevcc) & (\gevcc) & (\gevcc) \\ 
\hline
1     & 225  & 246  & 292 & 345  \\
\half & 204  & 233  & 282 & 337  \\
0     & 79  & 144  & 159 & 206  \\
\end{tabular}
\end{ruledtabular}
\end{table*}

\begin{acknowledgments}

We are grateful to M.~Kr\"amer for helpful discussions and for providing us with
detailed cross section information.  We also thank J.L.~Hewett and T.G.~Rizzo 
for many valuable discussions, and C.~Grosso-Pilcher for sharing results from
CDF with us and working on the combination of our limits.
Finally, 
we thank the staffs at Fermilab and collaborating institutions, 
and acknowledge support from the 
Department of Energy and National Science Foundation (USA),  
Commissariat  \` a L'Energie Atomique and 
CNRS/Institut National de Physique Nucl\'eaire et 
de Physique des Particules (France), 
Ministry for Science and Technology and Ministry for Atomic 
   Energy (Russia),
CAPES and CNPq (Brazil),
Departments of Atomic Energy and Science and Education (India),
Colciencias (Colombia),
CONACyT (Mexico),
Ministry of Education and KOSEF (Korea),
CONICET and UBACyT (Argentina),
The Foundation for Fundamental Research on Matter (The Netherlands),
PPARC (United Kingdom),
Ministry of Education (Czech Republic),
and the A.P.~Sloan Foundation.
\end{acknowledgments}

% LIST_OF_VISITOR_ADDRESSES.TEX                            5/18/01
%
%\bibitem[*]{lehner}
%Visitor from University of Zurich, Zurich, Switzerland.
%
\vskip0.2cm

\noindent\hrulefill

\vskip0.2cm

\noindent $^*$ Visitor from University of Zurich, Zurich, Switzerland.

\bibliography{lq_prd}

\begin{thebibliography}{10}
\bibitem{eta_def}
\bibinfo{note}{Pseudorapidity is defined as $\eta = -\ln \tan
  {{\theta}\over{2}}$, where $\theta$ is the polar angle measured with respect
  to the proton beam and relative to the $z$ position of the interaction
  vertex. We also use a variable called ``detector pseudorapidity,''
  $\eta_{\text{det}}$, for which $\theta$ is measured relative to the center of
  the detector, $z=0$.}
\end{thebibliography}

\end{document}